\documentclass[a4paper,twoside]{article}
\newcommand{\affil}[1]{$^{\rm #1}$}
\usepackage[authoryear]{natbib}
\bibpunct{(}{)}{;}{a}{}{,}
\usepackage{graphicx}
\usepackage{amsmath}
\usepackage{amssymb}
\usepackage{comment}
\usepackage{color}
\date{} %Please leave the date blank
\title{\large\bf\flushleft The Supershell--Molecular Cloud Connection: \\Large-Scale Stellar Feedback and the Formation of the Molecular ISM}
\author{\parbox{\textwidth}{\flushleft
\vspace{-0.5cm}
{\it J. R. Dawson\affil{A,B}}\\
\vspace{0.4cm}
{\small \affil{A}\,School of Mathematics and Physics, University of Tasmania, Sandy Bay Campus, Churchill Avenue, Sandy Bay, TAS 7005}\\
{\small \affil{B}\,Email: joanne.dawson@utas.edu.au}}}
\begin{document}
\twocolumn[
\begin{changemargin}{.8cm}{.5cm}
\begin{minipage}{.9\textwidth}
\vspace{-1cm}
\maketitle
%
%
%%%%%%%%%%%%%     ABSTRACT    %%%%%%%%%%%%%
%Abstract of no more than 200 words here.
\small{\bf Abstract:}
%We review the role of 
%theoretical background and observational evidence for the formation of molecular clouds via 
%large-scale stellar feedback in the formation of molecular clouds. 
The accumulation, compression and cooling of the ambient interstellar medium (ISM) in large-scale flows powered by OB cluster feedback can drive the production of dense molecular clouds. %, approaching the problem from the point of view of the evolution of the neutral ISM.  
We review the current state of the field, with a strong focus on the explicit modelling and observation of the neutral interstellar medium. 
Magneto-hydrodynamic simulations of colliding ISM flows provide a strong theoretical framework in which to view feedback-driven cloud formation, as do models of the gravitational fragmentation of expanding shells. Rapid theoretical developments are accompanied by growing body of observational work that provides good evidence for the formation of molecular gas via stellar feedback -- both in the Milky Way and the Large Magellanic Cloud. % measurements of the relative importance of this process on galactic scales are just beginning, but remain a challenge for future work. 
The importance of stellar feedback compared to other major astrophysical drivers of dense gas formation remains to be investigated further, and will be an important target for future work. 
%theoretical basis and observational evidence for molecular cloud formation via this process, with a strong focus on . 

%of this process compared to other astrophysical drivers are still challenging, however. Future work 

%We discuss the current state of both observations and theory, and briefly outline the ... directions for future development.  %The mechanical energy input from OB clusters drives large-scale flows that can sweep up and compress the ambient ISM to trigger episodes of molecular cloud and star formation. We review the theory underpinning this process
%This is the phenomenon (we review the theory?). We review the obs literature. There is good observational evidence that this process occurs [both in MW and elsewhere], but little indication of its importance compared to other drivers of molecular cloud formation. OR NOT. I DON'T KNOW.

%%%%%%%%%%%%%     KEYWORDS    %%%%%%%%%%%%%
\medskip{\bf Keywords:} galaxies: ISM, ISM: bubbles, ISM: clouds, ISM: evolution, ISM: structure, stars: formation
% Please write all keywords in lower case. PASA uses the
% standard list of subject headings adopted by The Astrophysical Journal
% and available from http://www.journals.uchicago.edu/ApJ/keywords_text.html.
% Keywords are separated by em-dashes, i.e. ---

%%%%%%%%DO NOT EDIT%%%%%%%%%%%%
\medskip
\medskip
\end{minipage}
\end{changemargin}
]
\small
%%%%%%%%EDIT FROM HERE%%%%%%%%%%%%

\section{Introduction}

%\textbf{Editing needed. Molecular gas vs star-forming gas} 
The formation of dense, star-forming gas %from the warm atomic interstellar medium (ISM) 
from the diffuse interstellar medium (ISM) is a key process in the evolution of galaxies, and a major unsolved problem in astrophysics. While the accumulation, cooling and fragmentation of the ISM into self-gravitating, star-forming structures has been extensively studied -- particularly from a theoretical perspective -- the subject remains a challenging one; involving the interplay between a complex array of physical, chemical and astrophysical processes. Many key questions remain unanswered, including which astrophysical drivers are responsible for the majority of dense gas formation, the timescales of cloud formation and destruction, and how the details of the fragmentation process set the basic foundations for subsequent star formation. %lifecycle of the ISM. %, %ISM evolution, %of the interstellar medium (ISM), 
In the present-day, metal-rich universe, the ISM in star forming clouds is almost always in the molecular phase. %, since the requirements for dense gas formation are very similar to the requirements for molecule formation \citep{glover12}. %High column densities are essential in shielding the material from the background UV radiation field, 
%Whether or not the ISM is in its molecular phase is therefore a primary diagnostic in identifying star-forming gas, and 
The key question can therefore be phrased as: How, when and where do we form molecular clouds? 

%Molecular clouds are high-density, high column-density entities. \textbf{where cooling is efficient and where gas is well-shielded from UV radiation. Cold, dense gas can form without the need for molecular cooling, and star formation can technically proceed without gas becoming molecular, but since the conditions for molecule formation are essentially identical to the conditions for cold, dense gas formation, molecular gas is synonymous with star forming gas in all but the most metal-poor systems \citep{glover12,clark12}} The former condition is essential for cooling, and for molecule formation to proceed on reasonable timescales, and the latter ensures that molecular species are shielded from dissociating UV radiation. % the shielding 
The basic requirements for molecular cloud formation are high densities and high column densities. % are the critical requirements for molecular cloud formation. 
The former is essential for effective cooling (and also facilitates rapid molecule formation) and the latter is necessary to shield the gas from UV heating (and also prevents the photodissociation of molecular species). Molecular cloud formation therefore typically involves large-scale compressive events capable of piling large quantities of material into small volumes. 
% the first stage of the star formation process, and sets fundamental boundaries on star formation rates. %The locations, mechanisms, timescales and drivers of molecular cloud formation %determine the initial conditions of -- and 
%determine 
%are what ultimately determine where and when star formation can occur, 
%set fundamental constraints on the star formation activity in galaxies. % is  fundamental boundaries on a star formation rates, and it is therefore vital to know what
Astrophysical drivers %of molecular cloud formation 
include global gravitational instabilities in galaxy disks \citep[e.g.][]{wada00,kim02,tasker09,bournaud10}, the accumulation of matter in spiral shocks \citep[e.g.][]{kim06,dobbs06,dobbs08}, and compression in expanding shells driven by stellar feedback \citep[e.g.][]{mccray87, hartmann01,ntormousi11}; all aided by turbulence that acts to enhance density on a range of scales \citep[e.g.][]{elmegreen02b,maclow04,glover07,mckee07,federrath12}. %However, the primary drivers -- as well as the details of the physics -- are still very much open to debate. 

This review will examine the role of large-scale stellar feedback in the formation and evolution of molecular clouds. % -- %This formation mechanism is of particular interest, since it is 
%a key part of how star formation self-regulates. % in galaxies. 
%On local scales stellar feedback is both a disruptive to parent molecular clouds and a regulating influence on star formation, since it heats and ionises the gas, injects turbulence to support against further collapse, and drives material out of the cloud itself \citep[e.g.][]{dale12}. %which are ultimately driven apart by the stars they form. %Simulations generally find that the injection of ionising radiation and mechanical feedback into parent clouds ultimately inhibits global collapse and acts to maintain low star formation efficiencies \citep{joung06,dale12}. %, although some triggering may occur .
%However, on larger scales t
This route to molecular cloud formation is of particular interest, since it is key element of how star formation self-regulates. 
The cumulative energy input from multiple stellar winds and supernovae may form new molecular gas through the sweep up and sustained compression of the ambient atomic medium in giant `supershells' around OB clusters. %Multiple stellar winds and supernovae from OB clusters drive sustained shock fronts into the ambient medium, sweeping it up into giant supershells, and persisting over large enough distances and long enough timescales that sufficient material can be accumulated to form new molecular clouds \citep{mccray87}.  %While the intense radiation and mechanical feedback from massive stars are disruptive to their parent clouds, %dissipate parent clouds, 
%the cumulative energy input from stellar winds and supernovae may drive molecular gas formation, %have the opposite effect on larger scales; forming new molecular gas 
%through the formation of new molecular clouds from the material accumulated in giant supershells around OB clusters \citep{mccray87}. %From a theoretical perspective, r
Rapid cooling of this swept-up gas -- in combination with gravitational, fluid dynamical and thermal instabilities -- results in the fragmentation of supershell walls into dense clouds, which will become molecular (and eventually star-forming) where density and column density requirements are met. %While more broadly, 
%In modern theory %supershells fall within the scope of systems covered by the 
This process is often discussed in the context of %paradigm of 
the flow-driven model of molecular cloud formation, which envisions dense gas and star formation as a rapid, dynamical process, and molecular clouds as short-lived entities that form at the interfaces of turbulent flows %, of which supershells are one potential driver 
\citep[see][]{walder96,hennebelle99,elmegreen00,hartmann01,audit05,vazquez06,vazquez07,hennebelle08,heitsch08c,banerjee09,inoue12,clark12}. % rapidly at the %pressure-confined 
%interfaces of turbulent ISM flows \citep[see review by][]{vazquez10}. 
%Large-scale stellar feedback is indeed one means of driving these flows \citep{hartmann01}. 
Magneto-hydrodynamical models of this process have been developing rapidly in the last decade. 

Despite theoretical advances, the role of large-scale stellar feedback in molecular cloud formation %-- either in the Milky Way or elsewhere -- 
remains an outstanding question. 
%While much evidence suggests that %the net effect of 
%stellar feedback generally inhibits or regulates star formation both on individual cloud and galactic scales \citep[see e.g.][]{joung06,hopkins12,dale12}, this does not preclude the initial formation of some fraction of the molecular cloud population through the same feedback processes. 
It is clear that the feedback from massive stars plays a key role in both structuring and supporting the disk ISM \citep[e.g.][]{avillez01,joung09,dobbs11,hill12,hopkins12}, and is likely %by stellar clusters and isolated OB stars 
responsible for triggering a significant fraction of observed star formation on local scales in existing molecular clouds \citep[e.g.][]{boss95,yamaguchi99,hosokawa06,dale07,leao09,deharveng10}. However, its role in the initial production of dense gas is not well constrained. There is evidence to suggest that the formation of dense clouds and clumps on galactic scales is primarily driven by a combination of global gravitational instability and turbulence, %is what dominates 
%that dominates in the initial production of dense (molecular) clumps and clouds %on galactic scales
\citep[e.g.][]{wada01,elmegreen02b,tasker09,bournaud10,maclow12}, with stellar feedback relegated to a secondary role. % in further processing this material.  
%Indeed, while galactic-scale gravity is implicated in the early stages of dense cloud formation in galaxies, supernova 
%Nevertheless, it is expected to drive more localised episodes of compression and cloud formation, and there is a need to constrain the relative importance of this process, both in the Milky Way and elsewhere. 
%These various theories await observational confirmation. 
In this picture global self gravity produces large dense gas complexes and spiral arms, while the localised energy input from OB clusters both triggers star formation in existing clouds, and drives the formation of some extra molecular material \citep{elmegreen02b}. %, with the exact level of its contribution determined by the level of  feedback and the characteristics of the galaxy \citep{dobbs12}.
The magnitude of the stellar feedback contribution, however, remains to be confirmed. 
Observationally, there is a large body of literature that makes at least some reference to the formation, evolution and destruction of molecular clouds in relation to feedback supershells or superbubbles %(the term is used largely interchangeably), 
\citep[e.g.][]{jung96,patel98,fukui99,kim00,matsunaga01,yamaguchi01a,yamaguchi01b,dawson08a,dawson11b,dawson11a}. 
However, much of this work consists of case-studies of individual objects, in which it is difficult to conclusively demonstrate triggered cloud formation, and discussion on the origin of the molecular gas is often speculatory. %there is no consensus on the magnitude of the contribution of large-scale stellar feedback to molecular cloud formation -- either in the Milky Way or elsewhere. 
%work has lagged behind theory, in the sense that while 
 %convincingly demonstrating formation %rather than mere association 
%often proves challenging. %, and has rarely been attempted in a robust or quantitative way. 
%Indeed, there is as yet no consensus on the magnitude of the contribution of supershells to molecular cloud formation rates -- either in the Milky Way or elsewhere. 
%The issue is further complicated by %the fact that the passage of a supershell shock front through the ISM can be destructive to 
%While 
%a growing body of 
%numerous studies document molecular clouds %are often observed to be 
%closely associated with stellar feedback superstructures, formation from the swept-up medium is rarely the only plausible origin for any given object. 
One challenge is that the real ISM is structurally very complex. It is characterised by a frothy structure of loops, filaments, carved out tunnels and irregular interlocking shells and bubbles, exhibits fractal structure across all size scales, and is peppered with molecular clouds whose individual evolutionary histories are ambiguous. The interaction of supershell shock fronts with pre-existing dense gas is also common, and further complicates the interpretation of associated clouds. %and can disrupt, entrain, compress or eventually destroy a molecular cloud, potentially triggering star formation in its wake. 
Until recently, the limitations of molecular line observations meant that most work on the molecular ISM in supershells was carried out in the Milky Way, where line-of-sight confusion exacerbates observational difficulties. %These observational challenges contribute to a situation in which there is still no consensus on the magnitude of the contribution of large-scale stellar feedback to molecular cloud formation -- either in the Milky Way or elsewhere. %,  on  that the contribution of stellar-feedback to molecular cloud formation rates may be small \citep{joung06,dobbs11}, 
%and there is a strong need for data to constrain theory. 
It has therefore proved challenging to address the role played by large-scale stellar feedback to molecular cloud in a quantitative sense, or examine its importance in entire galactic systems. %, and much work has been phenomenological in nature. % -- either in the Milky Way or elsewhere. \
Nevertheless, when taken as a whole the literature provides a good case for molecular cloud formation in supershell walls, and there is now a growing body of work approaching the problem from a more quantitative or statistical standpoint, %carefully examining the relationship between supershells and the molecular ISM, 
spurred on by ever-improving observational capabilities 
\citep[e.g.][]{matsunaga01,yamaguchi01b,book09,dawson08a,dawson11a}. %, particularly in molecular line observations.  %\textbf{[Need something more here]}

This paper reviews the theoretical basis and observational evidence for feedback-driven molecular cloud formation, focussing primarily on the evolution of the neutral ISM and the relationship between supershells %as the defining signatures of large-scale feedback, 
and the molecular medium. %between these deterministic structures and the molecular ISM. 
In Section 2 we provide some background %in section \ref{} 
%classic literature on supershells and superbubbles, summarising the key aspects of their 
on the feedback-structured ISM, the evolution of a classical shell, and a brief overview of the ways in which supershells may influence the molecular ISM. %(\S\ref{bubblebath}), the classical picture of the formation and evolution of supershells in disk galaxies (\S\ref{idealisedshell}, \S\ref{blowout}), and the ways in which they may influence the molecular medium (\S\ref{molecularism}). % and the early observational work that drove it. %and tying it to our modern understanding of the  as well as some basic observational characteristics. 
Section \ref{theory} deals with the theory and modelling of molecular cloud formation, including the gravitational fragmentation of expanding shells, the colliding flows model of molecular cloud formation, and numerical simulations of the ISM in galaxy disks. %in supershells, 
%beginning with a brief overview of molecule formation and destruction (\S\ref{molecules}), then describing the gravitational fragmentation of expanding shells (\S\ref{gravity}), the applications of the flow-driven model of molecular cloud formation to supershell systems (\S\ref{flows}), and large-scale models of the impact of stellar feedback on galaxy disks (\S\ref{wholedisk}).  %and its application to supershell systems, and finishing with large-scale simulations of of stellar feedback in galaxy disks. %moving on to whole-disk-scale simulations of stellar feedback in galaxy disks, and ending with the findings of the flow-driven model of molecular cloud formation and its application to supershell systems (section \ref{theory}). %We briefly touch on the subject of the shock-disruption of dense clouds before 
In section \ref{preexisting} we touch briefly on the relevant theory pertaining to the interaction of a supershell shock front with existing dense clouds, although the complex subject of triggered star formation is given only a brief mention. 
%We also give an overview of some relevant aspects of the interaction of existing molecular gas with supershell shocks in section \ref{preexisting}, including the destruction (\S\ref{destruction}) and induced collapse (\S\ref{triggering}) of dense clouds. % comments on the destruction of per-existing molecular clouds in shell-cloud interactions ... ... ... 
Section \ref{observations} describes observations of molecular gas in feedback superstructures, working outwards from the local ISM, to the Milky Way as a whole, to the Magellanic Clouds, and  %beginning with  (\S\ref{obsgeneral}) in the local ISM (\S\ref{obslocal}), the Milky Way as a whole (\S\ref{obsmw}), and the Magellanic Clouds (\S\ref{obslmc}), \
presenting strong evidence that molecular cloud formation via large-scale stellar feedback is occurring both in the Galaxy and in the LMC. %, including early attempts to measure the relative importance of this process. 
Finally, section \ref{summary} provides a brief summary and outlines some directions for future development. %, and outlining early attempts to measure its relative contribution to molecular cloud formation rates. We discuss outstanding questions and future directions and summarise in section \ref{summary}.

% It is preferable to embed your figures in the text as in the following example
\begin{figure*}[t]
\begin{center}
\includegraphics[scale=0.7, angle=0]{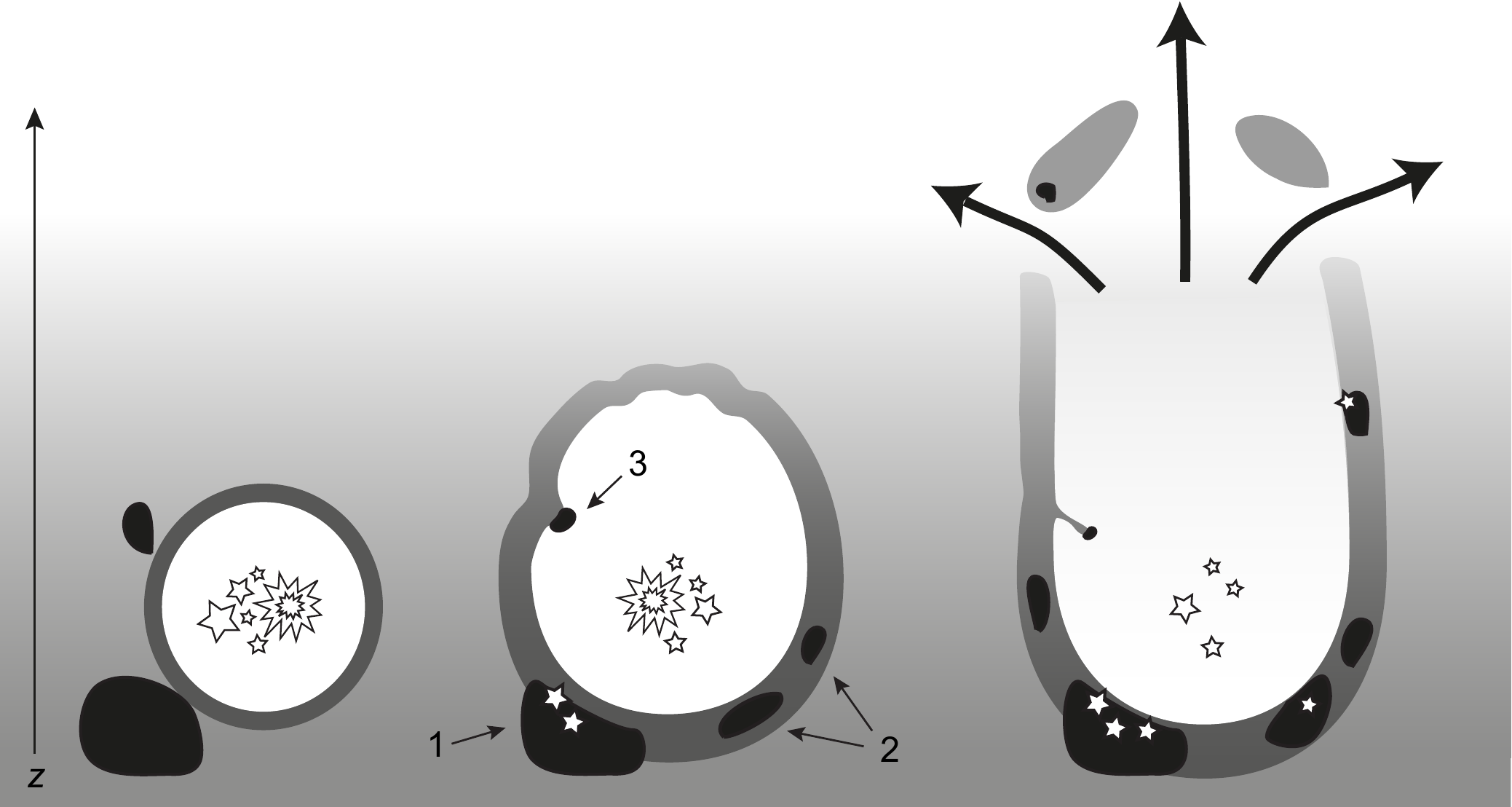}
\caption{\small Cartoon showing an edge-on view of the evolution of a supershell in the Galactic Plane (time sequence from left to right), illustrating some of the ways in which large-scale stellar feedback can affect the molecular ISM. Here black clouds represent molecular gas and the greyscale is the ambient atomic ISM. Labels show examples of 1. the triggering star formation in existing molecular gas, 2. the formation of new molecular clouds, 3. the disruption and entraining of existing molecular clouds. }
\label{cartoon}
\end{center}
\end{figure*}

\section{Background}
\label{basics}

\subsection{The Galactic Bubble Bath}
\label{bubblebath}

%It has long been known that t
The energy input from massive stars significantly impacts the structure and evolution of the interstellar medium. The ISM of star-forming galaxies is riddled with the footprints of this stellar feedback, in the form of cool shells, hot bubbles and evacuated cavities seen in multiple tracers, %and in multiple wavelengths, 
from X-ray to radio. The existence of such structures in the ISM has been extensively documented throughout the last century \citep[see review by][]{tenorio88}, and led \citet{brand75} to coin the phrase `the cosmic bubble bath' to describe the structure of the Galactic Disk. The volume filling factor of feedback bubbles in the Milky Way is thought to lie somewhere between $\sim0.05$--1.0, with most estimates lying at the lower end of this range \citep{oey97,oey01,mcclure02,oey04,ehlerova05}.

The term `supershell' was first used by \citet{heiles79} to describe atomic hydrogen shells with estimated formation energies of $E\gtrsim10^{52}$ erg, but is now generally used loosely to refer to any large ($R\gtrsim100$ pc) shell-like structure. % assumed to be formed by multiple stellar sources. 
The theory of the formation of these superstructures via correlated supernovae and stellar winds was developed during the following years \citep[e.g.][]{mccray79,bruhweiler80,tomisaka81,maclow88}, and although alternative formation mechanisms have been proposed \citep[e.g.][]{tenorio81,loeb98,wada00}, stellar feedback is successful in accounting for the observational characteristics, population-density and energetics of the majority of objects \citep[e.g.][]{mccray87,ehlerova96,mcclure02,weisz09,warren11}. 
To the present day, a steady stream of observational work continues to catalogue and study new supershells both in the Milky Way and in external galaxies \citep[see e.g.][]{kim99,mcclure02,ehlerova05,bagetakos11}. %\textbf{TIMESCALES. Probably need this somewhere since a lot of the issues centre around whether you can form MCs before shell dies. Though the density.}

\subsection{Evolution of a Classical Supershell}
\label{idealisedshell}

%The evolution and structure of a supernova remnant expanding into a uniform ISM was described analytically by \citet{cox72}, building on the work of many past authors \citep[e.g.][]{sedov59,heiles64,shklovskii68}, and an equivalent treatment for stellar wind bubbles was developed soon after by \citet{castor75}. 
The classical analytical model for a stellar wind bubble expanding into a uniform medium was derived by \citet{weaver77}, and modified for a system formed by multiple stellar winds and supernovae by \citet{mccray87}. %, and provides a good basis from which to describe the basic structure and formation of a classical ISM bubble. 
The early evolutionary stages of the system are dominated by stellar winds, which create a hot, low-density cavity into which subsequent supernovae inject their energy. By $t\sim5\times10^6$ yr these winds have switched off, and supernovae continue to inject energy until $t\sim5\times10^7$ yr, with the mass in the cavity acting as a buffer to the supernova blast waves. %, so that their KE is converted to heat. 
During the stellar wind phase, the idealised bubble consists of an inner zone of hypersonic stellar wind, a shocked stellar wind layer, a shell of shocked ISM, and the ambient ISM. The system initially evolves adiabatically, expanding much faster than the radiative cooling timescales in any of the four zones, ending after several $10^3$ years, when radiative losses in the swept-up shell become important and it rapidly begins to cool. This drop in temperature is naturally accompanied by a drop in pressure, and the shell collapses to a thin, cool layer, compressed by the thermal pressure of the hot ($T\sim10^6$ K) interior. The expansion continues to be driven by %the pressure of the hot interior ($T\sim10^6$ K), 
this thermal pressure, with a growth rate of the radius of the shell, $R$, given by $R\propto t^{3/5}$. This relation holds until such time as radiative cooling of the interior gas begins to become important, estimated to occur at %a few $10^5$ yr for a single O-star driven stellar wind bubble, and -- They included it from here but the deviation was tiny!!
several $10^6$ yr for a typical cluster-driven superbubble \citep{mccray87}. A shell whose interior thermal energy has been entirely radiated away will expand by conservation of momentum alone, with $R\propto t^{1/4}$. The shell may contain an ionised inner skin for much of its life, with a thickness determined by the radiation output of the remaining central sources. 

%, and an outer layer of neutral material. In the classical model the ionised inner layer will remain at $T\sim8000$ K, while the temperature of the neutral shell will stabilise at $\sim 80$ K, assuming that it remains in the atomic phase. However, it has long been recognised that an initially atomic shell might accumulate sufficient material (and persist for long enough) to become molecular \citep[e.g.][]{hollenbach76,mccray87}, eventually fragmenting to form discrete molecular clouds and new stars \citep[see reviews by][]{tenorio88,elmegreen98}. This process is of particular relevance to supershells, which will quickly outgrow the remnants of their parent molecular clouds to sweep up a primarily atomic medium. We will return to the topic of supershell-triggered molecular cloud formation again in \S\ref{molshellstheory}. 

\subsection{Disk Blowout}
\label{blowout}

Shells with sufficient energy may expand rapidly along the Disk vertical density gradient, eventually breaking out entirely and venting their hot interior gas into the Halo. Accelerated vertical expansion leads to the growth of Rayleigh-Taylor (RT) instabilities, culminating in the break-up of the shell and the release of its interior gas \citep{maclow89,tenorio90}. The energy requirements for blowout depend sensitively on the structure of the local ISM into which the central cluster inputs its energy, as a well as on the distance of that cluster from the Galactic midplane, and the strength of the disk magnetic field \citep[e.g.][]{tomisaka92,tomisaka98}. % In an idealised smooth, stratified medium, the energy requirements for blowout correspond roughly to a stellar cluster containing $\gtrsim30$ supernova progenitors \citep[see][]{mccray87}, or higher if magnetic fields act to oppose vertical expansion \citep{tomisaka92,tomisaka98}. 
In a realistic, pre-structured medium, however, blowout almost certainly occurs relatively easily, as an expanding supershell seeks out low-density pathways through the inhomogeneous ISM \citep[e.g.][]{avillez01,avillez05,hill12}. These blown-out `chimney' systems form a vital link from the Disk to the Halo, supplying the latter with the energy and metal-enriched material %needed to explain many of its properties 
(\citealp{norman89}, \citealp[see also review by][]{dickey09}). %These `chimney' systems form a critical link in the cycle of energy and matter between the Disk and the Halo, and 
%The hydrodynamics of chimney formation have been investigated by a number of authors, taking advantage of %the %opportunities offered by 
%rapid developments in computational astrophysics that occurred in the late 80s and early 90s. 
%As a supershell's radius grows comparable to the disk scale height, the gas density gradient begins to become important. In a smooth, stratified atmosphere with scale height $H=\int_0^\infty{\frac{n(z)}{n_0}}\ dz$, expansion in the vertical direction will generally begin to accelerate within $\sim2H$, provided that the energy of the system is sufficient to prevent the shell stalling before this point (\citealp{maclow88}, see also \citealp{tomisaka86}). 
%This is typically found to occur for input luminosities of $\gtrsim10^{37}$ ergs s$^{-1}$, which corresponds roughly to a stellar cluster containing $\gtrsim30$ supernova progenitors \citep[see][]{mccray87}. 
%Magnetic fields parallel to the Plane may confine shells that otherwise would have blown out \citep{tomisaka92,tomisaka98}. On the other hand, when the assumption of a smooth atmosphere is removed and the ISM is replaced by a more realistic medium pre-structured by prior generations of feedback \citep[e.g.][]{avillez01}, blowout may occur more easily as an expanding supershell seeks out low-density pathways through the inhomogeneous ISM.

\subsection{%Supershells \& 
The Molecular ISM}
\label{molecularism}

The central question in this review is the role played by large-scale stellar feedback on the formation, destruction and distribution of the molecular ISM. Figure \ref{cartoon} shows a cartoon of a supershell expanding in a stratified medium containing pre-existing molecular clouds, which illustrates both the large-scale features of the system's evolution, and the range of effects it may have on the molecular ISM.
%The question of the effect of supershells on the molecular ISM is the central one of this review. %Upon the development of a theory of supershell formation via clustered feedback, i
An initially atomic shell may persist for long enough and accumulate sufficient material to become dense, cold and molecular, eventually fragmenting to form discrete molecular clouds and new stars. %\citep[e.g.][]{hollenbach76,mccray87}; 
%a process that will be discussed in detail in \S\ref{theory}. %We will spend some time examining this process in Section \ref{theory}. %The theory that has been developed process is discussed at length in Section \ref{theory}. %, below.  
A shell may also encounter existing dense clouds in its path, resulting in their dynamical disruption and eventual destruction. %of the cloud. % process.% (see Section \ref{destruction}). %potentially destructive nature of the 
%potentially destructive interaction of a supershell shock front with pre-existing molecular clouds; discussed briefly in Section \ref{destruction}. %A realistic supershell expands into a highly 
%In a realistic, inhomogeneous ISM, %medium, and
%The response of existing dense molecular structures to encounters with shells is a crucial part of determining the net effect large-scale stellar feedback has on the molecular medium. 
Such encounters are also quite capable of triggering star formation in pre-existing dense clouds, however, so that the formation and subsequent collapse of fresh molecular material is not the only route to star formation afforded by large-scale stellar feedback. While we do not focus on the question of triggered star formation in this review, it is prudent to point out that the `destruction' of a molecular cloud can not automatically be assumed to have a negative impact on star formation. %, and that \textbf{[dude, it's complicated.]}

\section{Molecular Cloud Formation in Supershells: Theory \& Modelling} %of Molecular Cloud Formation in Supershells}
\label{theory}

An extensive body of literature exists on the physics and chemistry of the neutral ISM, and on the transition between its atomic and molecular phases. Here we focus on that work which is of particular relevance to the formation of molecular clouds by large-scale stellar feedback -- either through direct modelling of supershell systems, or through the development of general theory of ISM flows that includes those driven by supernovae and stellar winds. 

\subsection{Molecule Formation \& Destruction}
\label{molecules}

While recent work argues that the presence of molecules is not necessary for the formation of cold, dense gas \citep{glover12,maclow12}, molecular emission lines are the primary tracer of the star-forming ISM, and an understanding of how the major species form and evolve is essential in interpreting observational data. From a theoretical perspective, a region of the ISM in which hydrogen is predominantly in the form of H$_2$ constitutes a molecular cloud. However, observationally it is the abundance and properties of trace molecules -- in particular CO -- that determine whether a cloud is detectable. %Observationally, however, it is more often the abundance and properties of trace molecules that determine whether a molecular cloud is detectable. It is therefore common to consider the formation and destruction of both H$_2$ and CO -- the most common observational tracer. 
%However, from an observers point of view it is more often the abundance and properties of trace molecules that determine whether a molecular cloud is detectable
%In order to form molecular clouds, atoms must be converted into molecules and must remain in molecular form. 
%For the purposes of following the formation and destruction of H$_2$, m

Most treatments that explicitly follow molecule formation make use of approximate expressions for the formation rate of H$_2$ on dust grains \citep[e.g.][]{hollenbach71,tielens85}. %Following \citet{tielens85}, $R\approx6\times10^{-17}~(T/300)^{0.5}~n_{\mathrm{H}}n~S(T)$ cm$^{-3}$ s$^{-1}$, where $S(T)$ is a factor of order unity describing the ease with which H atoms bind to the grain surface. 
For conditions typical of the cold neutral medium \citep[CNM, $T\sim80$ K, $n\sim50$ cm$^{-3}$, see e.g.][]{field69,ferriere01}, the characteristic formation rate is $\sim3\times10^7$ yr, but a strong density dependence ensures that H$_2$ forms rapidly for densities that are even moderately enhanced with respect to canonical CNM values \citep[see also][]{glover11}. %Unlike H$_2$, efficient gas-phase reaction pathways are available for CO formation \citep{vandishoek88}. The detailed rate of CO formation depends on the solution of large networks of interdependent chemical reactions. However, models suggest that formation proceeds quickly once the conditions for shielding are met \citep[e.g.][]{bergin04}. 
For H$_2$ to survive, it must be shielded from dissociating photons in the energy range $11.08 < h\nu < 13.6$ eV. %but self-shields quite effectively \citep[e.g.]{draine96}. 
While full modelling of H$_2$ dissociation and shielding is complex \cite[see e.g.][]{vandishoeck88,draine96}, a combination of efficient self-shielding and dust shielding means that for UV field strengths within a factor of few of the Draine field, hydrogen will typically be in its molecular form for visual extinctions of $A_V\gtrsim0.1$--$0.5$ \citep[e.g.][]{franco86,vandishoeck88, bergin04,wolfire10}. This corresponds to column densities of roughly $\sim1\times10^{21}$ cm$^{-2}$ \citep{draine96}. 

For CO, the most commonly used line tracer of the molecular ISM, formation proceeds efficiently through collisional processes \citep{vandishoeck88}, but less efficient self-shielding means that abundances are primarily determined by visual extinction \citep{glover11}. The shielding requirements are therefore somewhat more stringent than for H$_2$, at $A_V\gtrsim0.6$--$1.0$ \citep{vandishoeck88, bergin04,wolfire10}. %are often used as an approximate minimum value necessary to sustain molecular hydrogen for a solar-neighbourhood-strength UV field \citep[e.g.][]{franco86,vandishoeck88,vandishoeck98}. %It should be noted that CO, the most common observational tracer of molecular clouds, shields less effectively than H$_2$, so that a significant fraction of `dark' molecular gas may exist, that is not recovered in 
%\textbf{Ooh, tenorio-tagle paper can go here for gathering up stuff}
%The implications of this include 
An interesting implication of this is the existence of a substantial quantity of `dark' molecular gas not seen in the usual surveys \citep[e.g.][]{grenier05,wolfire10}. %; a topic which we return to briefly in \S\ref{obsgeneral}. 

%In any case, both high densities and high column densities are a prerequisite for forming a molecular cloud. %, and some mechanism is needed that can accumulate and compress sufficient quantities of the ISM to achieve this state.

\subsection{Gravitational Instability of Expanding Shells}
\label{gravity}

For molecular clouds to fulfil their role as star formation sites, they must become self-gravitating -- or at least contain self-gravitating substructure. %The classical picture of molecular clouds as self-gravitating entities \citep[e.g.][]{solomon87,scoville87} has motivated 
%a lot of work on examining gravitational instability in shells.
Motivated by this, a number of authors have %examined the development of 
applied the theory of gravitational instabilities in expanding shells to %the context of 
the parameter space appropriate to supershells. % and large-scale propagating cloud/star formation. 
\citet{mccray87} use an analytical approximation for the fragmentation of an expanding shell into gravitationally bound clouds 
to derive a time for the onset of gravitational instabilities as $t\approx3.2\times10^7~N_*^{-1/8}~n_0^{-1/2}a_S^{5/8}$ yr, where $N_*$ is the number of supernova progenitors in the parent cluster, $n_0$ is the density of the ambient medium and $a_S$ is the magneto-sonic speed in the shell in km s$^{-1}$. For typical supershell parameters this suggests a timescale of a few $10^7$ years for shells to fragment into gravitationally bound clouds. The authors also state that they expect supershells to become predominantly molecular well before this, within timescales of $\sim10^6$ yr. However, this appears to be far too short for shells to accumulate sufficient material for H$_2$ shielding, and is in general not consistent with other models or with observations. 

A series of numerical models by \citet{ehlerova97}, \citet{efremov99}, \citet{ehlerova02} and \citet{elmegreen02a} use the thin shell approximation \citep{sedov59,kompaneets60} to study the gravitational fragmentation of supershells over a broad range of parameter space, with particular focus on galactic environment. It should be noted that recent work by \citet{dale09} and \citet{wunsch10} has demonstrated that the thin shell approximation deviates from the predictions of numerical simulations in cases where the external confining pressure is small. Nevertheless, it provides a useful and computationally cheap method of exploring the large regions of parameter space appropriate for supershells evolving under a wide range of different conditions.

\citet{ehlerova97} note that fragmentation timescales are insensitive to the initial energy input, but are strongly density dependent, and assume a realistic input cluster size of 40--100 supernova progenitors %Assuming an OB cluster of 40--100 supernova progenitors and a homogeneous medium, they find that systems with
to derive the condition that systems with $n_0\lesssim0.3$ cm$^{-3}$ never become unstable. Similarly, \citet{ehlerova02} derive a critical column density for gravitational fragmentation of $N\gtrsim10^{20}$--$10^{21}$ cm$^{-2}$ for realistic values of the energy input and sound speed in the ambient medium. Characteristic fragmentation timescales for solar neighbourhood densities of $n_0\sim1$ cm$^{-3}$ are long, at $\sim5\times10^7$ yr \citep{ehlerova97}, but would be significantly faster for moderately over-dense regions. Similarly, in a stratified medium, it is the dense central regions within the galactic plane that become unstable to gravitational collapse, while polar regions often remain stable throughout a shell's lifetime \citep[see also][]{mashchenko94}. Here, the thickness of the disk is critical in determining the fate of the system, %with shells %expanding in thin disks 
with Gaussian scaleheights of $\sigma\lesssim100$ pc resulting in shells that never fragment. Galactic rotation also strongly affects the evolution of the shell. Shear deforms shells into elongated ellipses, and mass accumulated in the shell slides to the tips, forming instability regions there. This phenomenon has also been noted by \citet{tenorio87}, who demonstrate that in a typical Milky Way environment these sites can accumulate sufficient matter to go molecular and form giant molecular clouds (GMCs). %, potentially forming giant molecular clouds (GMCs).   
The scenarios explored in these studies imply that the type of galaxy into which a supershell is born strongly influence its propensity for gravitational instability, and hence the likelihood of forming new molecular clouds through this mechanism. % and stars through this mechanism. 
\citet{elmegreen02a} derive a set of dimensionless conditions for this `triggering', and find that dwarf galaxies such as the LMC have an advantage over early type spirals such as the Milky Way.

%Recent studies by \citet{dale09,dale11} and \citet{wunsch10} have tested the thin shell approximation against full hydrodynamical treatments using both SPH and AMR codes, and find that the predictions of the thin shell approximation are

\begin{figure*}[t]
\begin{center}
\includegraphics[scale=0.3, angle=0]{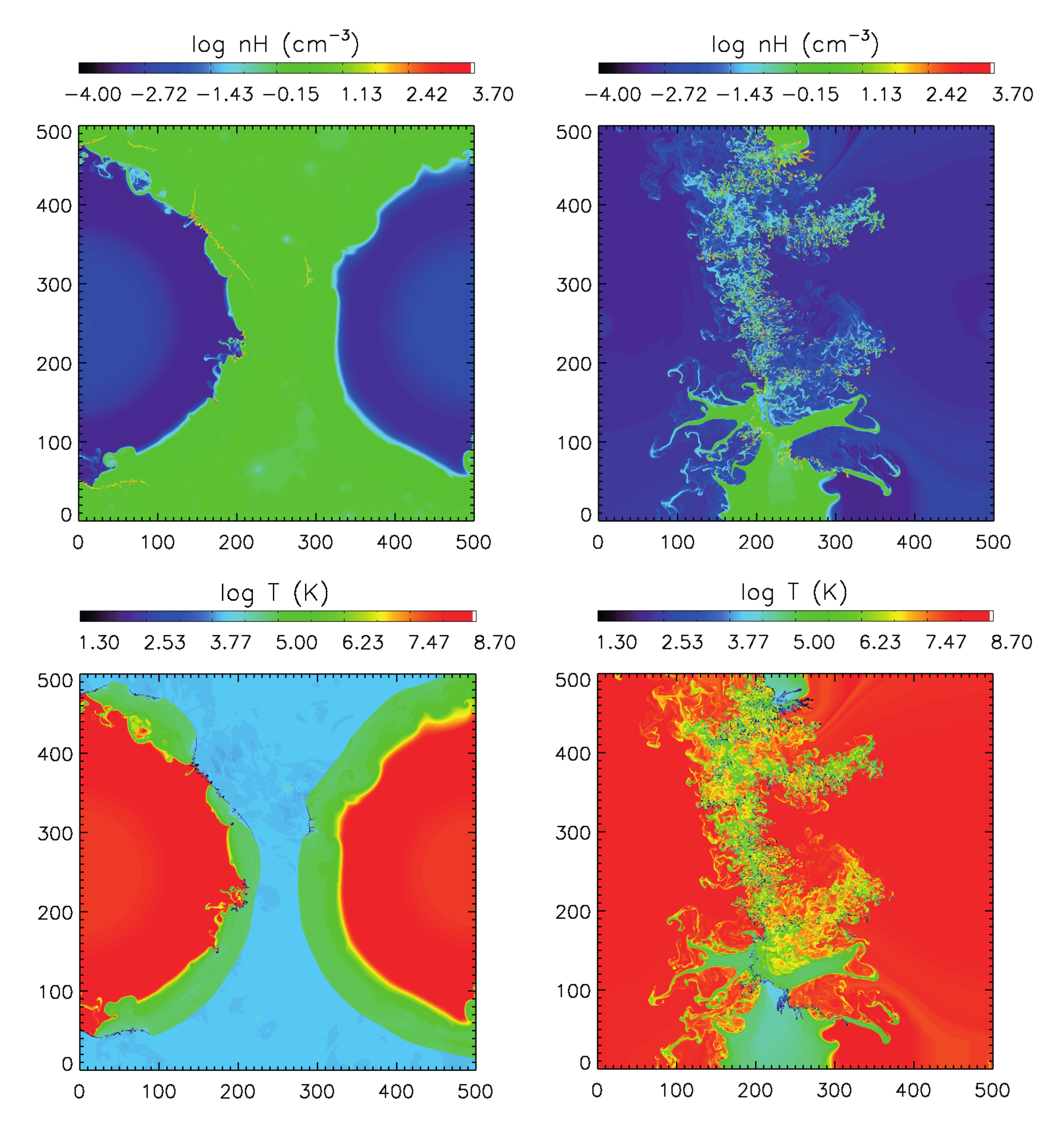}
\caption{\small High resolution, two-dimensional hydrodynamical simulations of cold gas formation at the interface of two superbubbles colliding in a turbulent diffuse medium (reproduced by permission of the AAS from \citealt{ntormousi11}). These snapshots show the evolution of the system at 3 Myr (left) and 7 Myr (right) after the start of the simulations. The top panels show hydrogen number density the bottom panels show gas temperature. After 7 Myr, copious amounts of clumpy and filamentary cold gas has been formed.}
\label{yeahntormousi}
\end{center}
\end{figure*}

%\textbf{write about self gravitating and grav fragmentation stuff: Can start with link from above paper. move on to mccray analysis, though note that their "fully molecular shell" prediction doesn't seem to allow sufficient columns and is not borne out by observations. Ehlerova and elmegreen treatments and what they mean. Then hartmann maybe? i.e. get these papers in then move onto colliding flows stuff?}% on some scale.

\subsection{Molecular Cloud Formation in Colliding Flows}
\label{flows}

%Approximations such as the thin shell approximation are an excellent way of reducing CPU load so that a problem can be examined over large regions of parameter space. However, the 
A realistic, turbulent ISM is subject to a range of processes that are not treated in the simplified gravitational stability analysis above, yet which may have a profound effect on its evolution. % from atomic to molecular gas. 
The last 10--15 years have seen a profusion of hydrodynamic and magneto-hydrodynamic (MHD) simulations that attempt to model the ISM in increasingly realistic ways, many focussing on the formation of molecular clouds. 

Much work centres around the paradigm of flow-driven molecular cloud formation, %fast molecular cloud evolution
in which the production of star-forming molecular gas is integrated into the modern picture of a dynamic, turbulent ISM.  In this picture, the compression, cooling and fragmentation of the atomic medium in colliding ISM flows produces clumpy sheets and filaments of cold, turbulent material, which go on to become highly-structured, self-gravitating molecular clouds once density and column density requirements are met 
\citep[see review by][]{vazquez10}. \citet{hartmann01} lay out part of the astrophysical motivation for this picture, which has its roots in %the need to explain 
observational evidence that molecular clouds must form rapidly, birth stars rapidly, and disperse rapidly \citep[see also][]{elmegreen00,hartmann03}. %, all within  
These authors use simple analytical approximations to argue that material accumulated in large-scale ISM flows will become molecular, magnetically supercritical and self-gravitating on roughly similar timescales, and that supershells formed by stellar feedback are excellent candidates for the drivers of such flows. %(Other possibilities include inflows from gravitational instabilities and large-scale turbulence.)

A combination of thermal and hydrodynamical instabilities, together with turbulent compression \citep[see e.g.][]{federrath10}, is critical in condensing cold gas from the warm ambient medium. %, and turbulence also plays a central role. %It has been recognised for many years that under the assumption of standard heating and cooling functions t
The atomic ISM is a thermally bistable medium \citep{field69,wolfire95}, with two linearly stable phases corresponding to the warm neutral medium (WNM) and cold neutral medium (CNM) -- the latter being the atomic precursor to molecular clouds. Trans-sonic converging flows of initially stable WNM %, which my be driven purely by turbulence, 
trigger runaway cooling and the formation of cold gas \citep{hennebelle99}, with dense structures developing rapidly on sub-parsec scales \citep[e.g.][]{heitsch06,vazquez06}.   %demonstrate that  \citep[see also][]{koyama00}. %\citet{koyama00} demonstrate a similar phenomenon in the case of strong shocks propagating into the WNM, in this case also following the chemical evolution of the ISM to form a thin layer of H$_2$ and CO molecules.
%\citet{heitsch08b} explicitly consider the role of different instabilities, and conclude that `dynamical instabilities disrupt large-scale coherent flows through generation of turbulence, while strong thermal fragmentation amplifies the resulting low-amplitude density perturbations, thus leading to small-scale, high-density fragments as seeds for local gravity to act upon.'
Models have progressed to include increasingly realistic physics and explore various different aspects of the cloud formation process, including the roles of turbulence \citep[][]{audit05,vazquez06,glover07}, magnetic fields \citep[][]{hennebelle08,inoue08,banerjee09,heitsch09,inoue09}, molecular chemistry \citep[][]{bergin04,glover11,clark12,glover12,inoue12} self-gravity \citep[][]{vazquez07,heitsch08a,heitsch08b,hennebelle08}, and the interplay between different instabilities \citep{heitsch08b}. 

Taken together, these simulations suggest that for expansion velocities and ambient densities typical of Galactic supershells ($v_{exp}\sim10$--20 km s$^{-1}$ and $n_0\sim1$--5 cm$^{-3}$), substantial quantities of CO-rich molecular gas can be produced on timescales of a few $10^6$ to $\sim10^7$ yr \citep{bergin04,heitsch08c,clark12,inoue12}, provided that magnetic fields do not provide significant support against collapse \citep{inoue08,inoue09,heitsch09}. %\citep{inoue09}. 
For the parameter space appropriate to the atomic ISM, cooling timescales are much shorter than the characteristic timescales for gravitational instability, %over all spatial scales, 
and thermal/dynamical instabilities dominate over gravity in driving fragmentation in the compressed medium \citep{heitsch08b}. % in the swept-up medium. %, and structure formation will occur first on the sub-parsec spatial scales where TI growth rates are fastest. 
This implies that small, dense, thermally-driven condensations will develop long before a supershell becomes gravitationally unstable, % on the longer wavelengths associated with gravitational instabilities, 
and that some molecular gas can form without the ISM becoming self-gravitating \citep{heitsch08c}. %the need for additional gravitational pressure. 
However, %while some molecule formation can occur before gravity becomes important,
global contraction helps to reduce overall timescales, and in particular may be important in accumulating sufficient material to shield CO so that clouds become observable \citep{heitsch08c}. %\citep{heitsch08a}. %At any rate, the onset of self-gravity and the formation of molecules are likely to occur on roughly similar timescales, and g
Once initiated, gravitational collapse will proceed hierarchically, beginning first in the dense substructures already imprinted on the medium by thermal instabilities and turbulence \citep{heitsch06,vazquez07}. 
One implication of this is that clouds are expected to begin forming stars rapidly once they become visible in molecular tracers, suggesting that the triggered formation of molecular clouds is synonymous with the triggered formation of stars. 

These results provide a strong theoretical basis for molecular cloud formation in supershells, and demonstrate its viability in systems with flow properties that are consistent with observed objects. The next stage is to explicitly simulate supershell systems as opposed to generalised flows. So far only one high-resolution hydrodynamical simulation has explicitly modelled flow-driven molecular cloud formation in a supershell system. \citet{ntormousi11} present 2D models of two superbubbles expanding into a uniform homogeneous or turbulent medium, and investigate the formation of dense gas around the bubble peripheries and interaction zone. Their supershells are blown by supernovae and time-dependent stellar winds with properties calculated from population synthesis models, and evolve with realistic flow velocities, timescales and size scales. Figure \ref{yeahntormousi} reproduces their snapshot of the temperature and density of the turbulent model at times of 3 and 7 Myr. By the end of a 7 Myr run, their simulation box is filled almost entirely by the shell systems, and a combination of non-linear thin shell, thermal and Kelvin-Helmholz (KH) instabilities has lead to the formation of copious amounts of clumpy and filiamentary cold ($T<100$ K) gas -- mostly associated with the shell collision zone. They find that $\sim65$--$85\%$ of the gas is contained in these small, dynamical structures, which have characteristic sizes of $\lesssim1$ pc and densities of $10^2$--$10^3$ cm$^{-3}$ -- fulfilling the approximate density and column density requirements for molecule formation.% and are dynamical entities that are constantly merging and splitting. % and ... in a warmer coronal bath of material. 

%These simulations focus largely on the interface between two bubbles, %which is the most efficient site of cold gas formation, 
%and it is clear that the development of instabilities proceeds less efficiently outside of the collision zone. 

The two-dimensional nature of these simulations imposes some limitations on the evolution and structure of the turbulence and fragmentation. Unlike in 3D models, where large-scale structures break up into smaller fragments, in 2D simulations they have a tendency to merge into larger agglomerations \citep[e.g.][]{federrath10}, potentially affecting the ease with which dense clouds are formed. Conversely, the lack of self-gravity in the simulations is expected to have the opposite effect -- making it harder to form dense clouds. The inclusion of magnetic fields in future models would also modify the properties of the collision zone \citep[e.g.][]{avillez05} and likely provide additional support. %Finally, it is also worth noting that the models focus largely on the interface between two bubbles, and it is clear that the development of instabilities proceeds less efficiently outside of the collision zone
Nevertheless, these models are a promising first step, and will be developed extensively over the coming years to include more realistic physics. %the extension of ... \textbf{blah blah blah encouraging models, look forward to more development.}

Finally, it is likely that both pre-existing inhomogeneities in the ISM and magnetic field orientation lead to a selection effect for molecular cloud formation in supershell walls. %, that may explain why observed shells do not generally contain \textit{more} molecular gas. 
%The cloud formation timescales implied for WNM densities of $n\sim1$ cm$^{-3}$ are generally of the order of a few $10^7$ yr, which, while not unfeasibly long for a supershell lifetime, are still longer than the estimated ages of many objects with a molecular cloud component (see \S\ref{observations}). 
%Cooling and cloud formation a strongly dependent on density, and even a
A moderate elevation in the mean density along a swept-up column significantly reduces cooling and molecule formation timescales, suggesting that cloud formation sites may be determined by the placement of pre-existing concentrations of moderately over-dense gas such as CNM sheets or filaments \citep[see also][for similar findings on galactic scales]{dobbs12}. Conversely, magnetic pressure has the power to completely prevent the formation of molecular gas when a significant component of the field exists perpendicular to the flow direction, %, magnetic pressure provides additional support against collapse, and the ISM may never reach the required conditions for molecule formation 
leading to the requirement that % This leads to a selection effect in which 
this perpendicular component %of the magnetic field perpendicular to the flow direction must 
be vanishingly small \citep{inoue09,heitsch09}. While it remains to be seen how the inclusion of self-gravity and non-ideal magneto-hydrodynamic processes affect the rigidity of this conclusion, the selection of cloud formation sites by field orientation and mean density may provide an attractive explanation for why real supershells are not \textit{more} molecular than they are.%the sparsity of molecular gas observed in real supershell systems.
%The problem of magnetic fields is a troubling one. ... The inhomogeneity of the initial medium further complicates the evolution of the ISM in real shells.
%In real shells expanding into a highly inhomogeneous and pre-structured medium, there are additional affects to be aware of. 
%A modest elevation in the mean density along a swept-up column %of gas 
%may significantly reduce cooling and molecule formation timescales, leading to a selection effect whereby cloud formation sites are determined by the placement of pre-existing concentrations of gas. 

%Ignoring the effects of magnetic fields, most models suggest that cold clouds with densities and column densities required for molecule formation will form from trans-sonic 

%For \citet{heitsch08b} perform 

\begin{figure*}[t]
\begin{center}
\includegraphics[scale=0.8, angle=0]{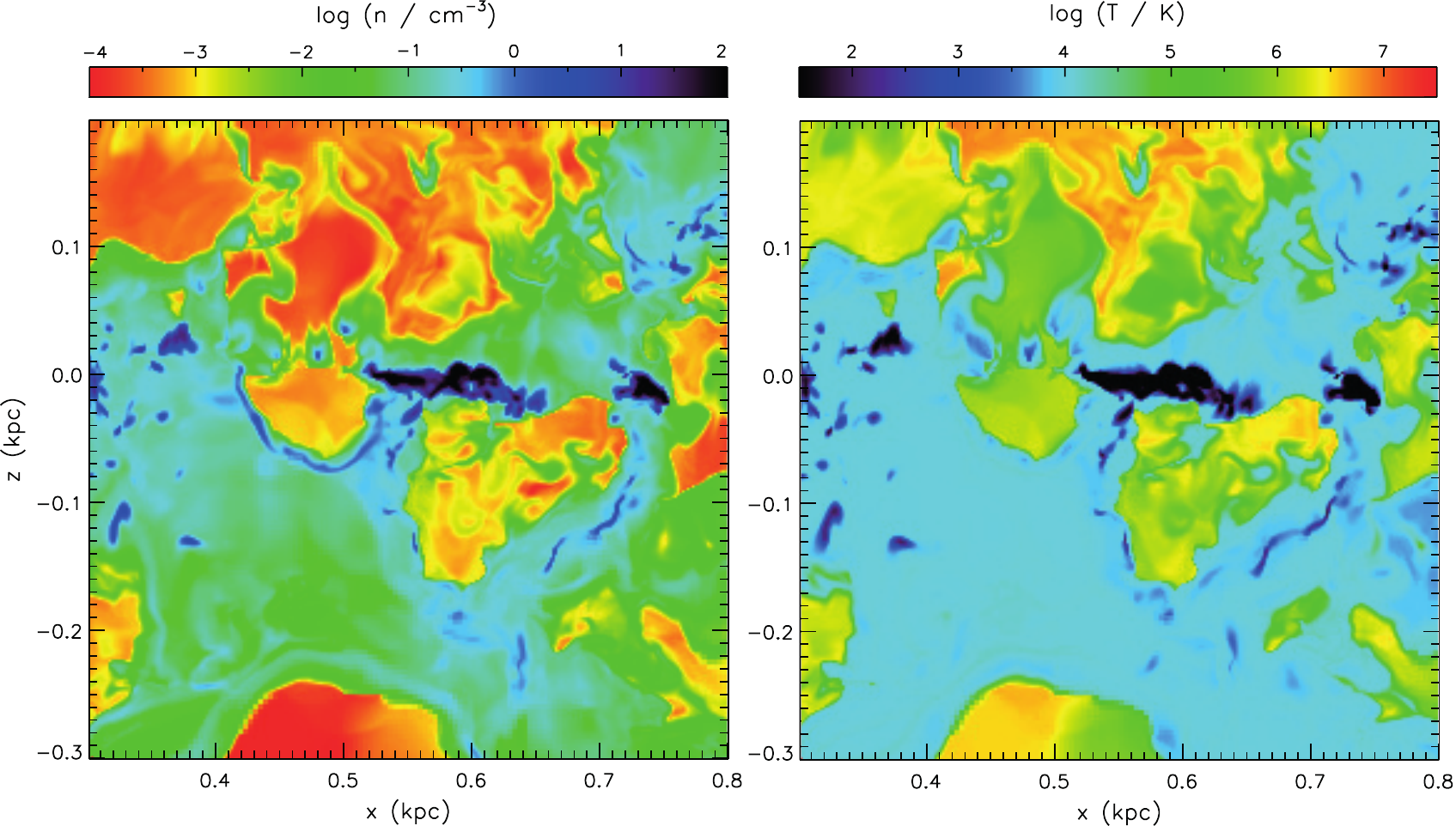}
\caption{\small Images of density and temperature from magnetohydrodynamic simulations in which supernovae drive turbulence and establish a multi-phase, stratified medium (reproduced with permission from the erratum to Hill et al, 2012). The images are two-dimensional slices through a three-dimensional simulation, where the x axis is in the midplane and the z axis shows the vertical distance from the midplane. The model used here is the magnetised model with $2 \textrm{ pc}$ resolution in the midplane (``{\tt bx50hr}'') described by \citet{hill12}. Energy injected by supernovae creates hot, low density remnants; surrounded by dense, long-lived filaments of cold gas.}
\label{alexfig}
\end{center}
\end{figure*}

%non-linear perturbations introduced solely by trans-sonic turbulence

% In the presence of external non-linear perturbations, 1D models by \citet{hennebelle99} demonstrate 

\subsection{Whole-Disk Models of the Feedback-Structured ISM}
\label{wholedisk}

While high-resolution numerical simulations have examined in detail the small-scale processes associated with gas cooling and fragmentation, another class of model tackles the global effects of stellar feedback on the structure and dynamics of the ISM. Such simulations generally model kpc-scale sections of galaxy disk at moderate resolution with realistic supernova input rates, and have been very successful in producing the cold, warm neutral, warm ionized, and hot components of the ISM with reasonable mass fractions, distributions and structure. %\citet{avillez01} model isolated, randomly distributed and clustered supernovae in a large section of galaxy disk, 
Salient features include a thin cold disk, a `frothy' disk of warmer material, and the existence of a great many shells, supershells and their broken fragments with `tunnels' channeling their hot interior gas out into the halo \citep[e.g.][]{avillez01,avillez05,joung06,joung09,hill12}. An example is shown in Figure \ref{alexfig}, from \citet{hill12}. %Of particular relevance to the issue of molecular cloud formation, 
While these models lack the resolution to properly resolve very dense gas formation, the fact that the bulk of the cold material is concentrated into high density clouds in shock-compressed layers around or between bubbles \citep[e.g.][]{avillez05} is highly consistent with the scenario described in \S\ref{flows}.

%These feedback-focussed models %focussed specifically on feedback %are an excellent proof of concept and 
%demonstrate the plausibility of the molecular cloud formation in supershell-driven flows, but  
%are unable to comment on the relative importance of this process in a galactic context. % importance relative to other drivers. %relative contribution of this formation mechanism when some of the other relevant astrophysics is included. 
The lack of self-gravity and realistic disk dynamics mean that while the above models are a promising demonstration of the feasibility of molecular cloud formation in stellar feedback flows, they are unable to address the critical question of its relative importance in a galactic context. Simulations of rotating, self-gravitating galaxy disks provide a more complex picture, in which large-scale gravitational instabilities and spiral arms play a major role in dense gas formation, and the contribution of stellar feedback %to dense gas formation
is by no means clear \citep[e.g.][]{wada01,shetty08,bournaud10,tasker11,dobbs11,dobbs12}. %However, progress is being made. 
Recently, \citet{dobbs12} have explicitly addressed the question of which mechanisms are responsible for driving the compressive motions that form the GMCs in their simulations, % for a range of different model galaxies, 
and find that the primary drivers depend on the properties of their model galaxies, such as the nature of the spiral potential and the level of star-formation feedback. While conclusions remain open, a future in which these predictions can be directly compared with observations is promising, as new facilities such as ALMA come online.

\section{Pre-Existing Molecular Clouds}
\label{preexisting}

%The theory of molecular cloud formation described above generally assumes a relatively smooth atomic medium. However, t
Supernovae and stellar winds propagate into a highly structured, inhomogeneous medium that already contains dense gas. %, and contains large quantities of dense gas. % with which an expanding shell must interact. 
%The ISM is highly structured and inhomogeneous, and can be thought of as pre-seeded with molecular clouds that a supershell will encounter as it expands. 
%the role of existing molecular clouds 
\begin{comment}
This fact should not be ignored when assessing the overall impact of stellar feedback on the molecular ISM. % between supershells and the molecular ISM.
%Clouds interacting with a stellar feedback flow may remain stable, may undergo triggered collapse and star formation, or may be completely shredded and destroyed. 
Mass loading from material stripped or evaporated from dense clouds also plays an important role in cooling shell interiors \citep[see e.g.][]{mckee77,cowie81,silich96}. It is therefore important to understand the factors that determine the eventual fate of a molecular cloud encountering a supershell, and the timescales on which the relevant physical processes occur.
%The eventual fate of this material -- stability, shock collapse and star formation, complete shredding and destruction -- and the associated timescales, 
\end{comment}
%Some key questions are: What is the eventual fate of a molecular cloud encountering a supershell? Will it collapse to form stars, remain stable or be completely destroyed? What are the timescales associated with these processes? %Under what conditions will a pre-existing cloud be accelerated and end up co-moving with the shell? 
The effects of cluster feedback on existing molecular clouds %will decrease as the system evolves, with the 
%ranges from the extreme to the mild. At one end there is the strong 
range from the violent ionisation, compression and dissipation of the parent cloud at early evolutionary stages, to the weak interaction of a cloud with an old, almost-stalled supershell. % at the end of its life. %The initial interaction with the parent cloud is an important aspect of how stellar feedback affects the molecular ISM, here %we do not talk explicitly about the physics of this here. 
%\textbf{[There's going to be general decreasing of destructive influence of cluster feedback as shell breaks out of parent cloud and gets larger and slower, with violent disruption of parent cloud at one end and weak effects of an almost-stalled supershell at the other. While initial destruction is important part of role of stellar feedback on molecular ISM, we do not talk explicitly about the physics of it here. This review is really about supershells. ]}
%Here we do not consider the interaction of young stellar sources on their parent molecular cloud -- this is assumed to have been dissipated early in the life of the cluster. 
The question of how feedback affects parent molecular clouds is a complex one, involving the combined effects of strong ionisation, radiation pressure and mechanical energy input on a pre-structured cloud. The interplay between these processes and the importance (or not) of triggered star formation is still a topic of much debate and will not be dealt with in this review \citep[see e.g.][and references within]{dale12,walch12}. Instead %we make the assumption that an OB cluster either destroys or blows apart its parent cloud relatively early in its evolution, and 
we focus primarily on the region of parameter space appropriate to a system that has expanded far beyond its birth site and out into the surrounding medium -- i.e. an object that would already be defined as a `supershell' by an observer. For an idealised shell of $R\gtrsim50$ pc this implies expansion velocities of no greater than a few tens of km s$^{-1}$ \citep{mccray87}; translating to shocks that are mildly supersonic in the WNM. % and strongly supersonic in the molecular ISM. %sound speeds in molecular clouds.
%\textbf{[Basically the issue isn't just what evolved shells are doing; it's the whole shebang, from beginning to end. That's why I don't feel comfortable with this whole approach. ]}

\subsection{Cloud Disruption}
\label{destruction}

The dynamical interaction of a dense cloud with a shocked flow is commonly parameterised in terms of the `cloud crushing time', $t_{cc}=(r_0/v_i)(n_{cl}/n_i)^{0.5}$, where $r_0$ is the cloud radius, $v_i$ is the velocity of the shock in the ambient medium, and $n_{cl}$ and $n_i$ are the number densities of the cloud and the ambient medium respectively. The simplest case is a steady planar strong shock and a smooth spherical cloud, in which radiative losses, thermal conduction, gravity and magnetic fields are neglected. 
Such clouds are significantly disrupted within a few $t_{cc}$, on a destruction timescale $t_{dest}$, defined as the time taken for the core mass of the cloud to reduce be a factor of $1/e$
%The time taken for RT and KH instabilities to disrupt an initially smooth spherical cloud so that its core mass is redcued by $1/e$ is 
%RT and KH instabilities will typically %accelerate and 
%destroy an initially smooth spherical cloud %via RT and KH instabilities 
%within 
%a few $t_{cc}$ 
\citep{klein94,nakamura06}. %Departure from the simplifying assumptions of the adiabatic case tends to lengthen destruction timescales. 
For the test case of a small ($r_0=2$ pc), moderately dense ($n_{cl}=200$ cm$^{-3}$) cloud encountering a gently expanding ($v_i=20$ km s$^{-1}$), thick, dense ($n_i=10$) shell, $t_{cc}\approx0.4$ Myr, implying $t_{dest}\lesssim2$--3 Myr. %Such clouds are also typically accelerated 

In reality this is likely a lower limit to cloud survival times. %a number of factors are likely to considerably lengthen this timescale. 
The Mach numbers of supershell shocks are relatively low ($\mathcal{M}\sim2.5$ for a 20 km s$^{-1}$ shock in the WNM), and timescales may be several times longer than the strong shock case \citep{nakamura06,pittard10}. %Low Mach number shocks 
Radiative cooling also cannot be ignored for supershell-molecular cloud interactions. Its inclusion %The inclusion of r
%Radiative cooling 
inhibits cloud destruction, encourages the formation of over-dense clumps and filaments and can significantly extend timescales above the adiabatic limit \citep{mellema02,fragile04,orlando05}. The role of magnetic fields is more complex -- while they tend to inhibit the development of hydrodynamic instabilities and reduce mixing, both orientation and field strength are important, and some configurations may result in more efficient fragmentation of the cloud material \citep{gregori99,orlando08,shin08}. Conversely, environmental turbulence has been shown to speed up cloud destruction, though the magnitude of the effect is sensitive to the properties of the assumed turbulence and is still under investigation \citep{pittard09,pittard10}. %Conversely, the inclusion of thermal conduction is found to inhibit the formation of disruptive KH and RT instabilities and extend survival times \citep{orlando08}. 
Nevertheless, it seems reasonable to assume that dynamical disruption of molecular clouds by an evolved shell proceeds reasonably slowly, on timescales comparable to the shell lifetime. %it seems reasonable to assume that dynamical disruption by an evolved shell will not destroy molecular clouds will generally survive dynamical disruption by an evolved shell. %, and will remain observable for timescales comparable to the shell lifetime.
It is worth noting, however, that the physical stripping and dissipation of cloud material also renders a disrupted cloud increasingly vulnerable to the UV dissociation of CO. The observable lifetime of the entity we see as a CO cloud may therefore be shorter than the survival time of the dense material itself \citep[see also][]{dawson11a}.

The efficiency of momentum transfer is also important, because it determines how readily the shocked cloud can be accelerated by the expanding shell. %, and therefore informs the interpretation of features in shell walls. %For adiabatic models, the drag time (%The drag time is typically less than that cloud destruction time by a factor of 2$Most models find that shocked cloud material is accelerated to $\sim70\%$ of the post-shock velocity  
Adiabatic models find that characteristic drag timescales are generally comparable or lower than $t_{dest}$, suggesting that a shocked cloud -- or its remaining fragments -- will undergo significant acceleration before being completely disrupted \citep{klein94,nakamura06}. Higher density contrasts and lower Mach numbers result in less efficient acceleration \citep[e.g.][]{pittard10}, as does any process that reduces the surface area of the cloud perpendicular to the shock direction \citep[e.g. radiative cooling;][]{orlando08}, while environmental turbulence and certain magnetic field orientations may increase it \citep{shin08,pittard09}. As in the case of cloud destruction, the details of the cloud acceleration depend both on the model physics and on the cloud-shock parameters, and the final velocity of the accelerated cloud material may be anywhere between $\sim0.1$--0.8 times the initial shock velocity. %However, acceleration of the cloud material to a  initial shock speed is generally plausible. % before a cloud is completely destroyed.  
%is defined as the time for the cloud to be accelerated to a fraction of $1/e$ of the shock velocity, and in the adiabatic case is typically found to be between 0. times smaller than $t_{dest}$ \citep{nakamura06}. However, the acceleration of the cloud depends quite sensitively on the assumed parameters of the interaction. For the  
%While most accelerated cloud material will not reach the velocities of the initial shock, 
%\textbf{[Add something on drag. Basically it just has to be in there somewhere that while destroying this cloud material you'll also be moving it. While details of acceleration depend on included physics, as well as on cloud/shock parameters (particularly on density contrast), timescales are similar to destruction timescales, and final speeds are never 100\% of shock speed, so that you'll have some movement of existing cloud with shell, but Also on triggered star formation in more realistic clouds?]}

The majority of these studies assume the post-shock flow has an infinite depth -- a situation that is clearly inappropriate for the shell-cloud case. The finite depth of a swept-up shell limits the time of the flow-cloud interaction, and a cloud encountering a supershell will likely pass into the hot shell interior before it is destroyed, leaving a tail of shell material trailing behind it \citep{pittard11}. For a shell thickness of $\sim10$ pc, the example cloud described above will enter the interior on timescales of $\sim1$ to several Myr, depending on the drag efficiency. %, likely leaving a tail of shell material trailing behind it \citep{pittard11}. 
In the interior regime the material flowing past it will be more diffuse and less dynamically disruptive, but will likely be hot ionised medium. The classical evaporation timescale for clouds embedded in fully ionised medium is given by $t_{evap}\sim3.3\times10^{20}~n_{cl}~T_i^{-5/2}r_{cl,\mathrm{pc}}^2$ yr \citep{cowie77}, which equates to $\sim10^8$ yr for the example cloud, assuming $T_i=10^6$ K. This suggests that thermal evaporation is unlikely to be a dominant destructive influence. 

Finally, while we do not deal explicitly in this review with the %complex and involved
topic of triggered star formation, it should be stressed that the destruction of a molecular cloud can not be assumed to have a negative impact on star formation. Indeed, for moderate shock velocities appropriate to supershell systems, star formation may be triggered readily for a range of molecular cloud properties \citep[e.g.][]{vanhala98,melioli06,leao09}. These considerations are important in interpreting observations of supershell-associated molecular gas and stars.

\section{Observations of Molecular Gas in Supershells}
\label{observations}

\subsection{General Considerations}
\label{obsgeneral}

In this section we will review observational evidence for molecular cloud formation in supershells, as well as some broader examples of the interaction between stellar feedback and the molecular ISM. 

The most commonly-used spectral line tracers of the molecular ISM are the low-$J$ transitions of CO. % and its isotopologue $^{13}$CO. 
CO is the workhorse of molecular ISM astronomy, since its high abundance ($n_\mathrm{CO}/n_{\mathrm{H}_2}\sim10^{-4}$), low-lying rotational energy levels ($\Delta E/k\sim5$ K for $J$=1) and relatively low critical densities ($n\sim1000$ cm$^{-3}$ for the lower J transitions) mean that it is readily observable even in relatively diffuse, quiescent molecular gas. For a detailed physical and chemical census of a molecular cloud %-- right through from quiescent envelopes to hot, dense star-forming cores --  
it is possible to assemble a suite of lines from multiple tracers that probe a wide range of density and temperature regimes; %in order to probe different temperature, density and chemical regimes right through to hot, dense star-forming cores. This 
an approach that is particularly useful for detailed studies of the star formation process. However, from a molecular cloud formation perspective we are generally more concerned with identifying zones in which the ISM is simply in its molecular form. CO is an excellent choice for this, although it is worth noting that even CO fails to trace the most diffuse molecular gas, where hydrogen is in the form of H$_2$ but carbon is atomic \citep[e.g.][]{reach94,grenier05,wolfire10}.

%It is worth noting, however, that the shielding requirements for CO are more stringent than for H$_2$ (see also \S\ref{molecules}), meaning that there exists a substantial quantity of CO-dark molecular gas that is not seen in the usual surveys \citep{reach94,grenier05,wolfire10}. 

%, the proportion of which depends on the local configuration of the medium and radiation field \citep{grenier05,wolfire10}. 
%This is of particular concern when searching for transition-state material, in which hydrogen is in the form of H$_2$ but CO abundances may still be low. IR emission from dust grains, which probes total gas column density, can provide insight into this component \citep[e.g.][]{reach94,leroy11}, as can UV absorption studies against background sources \citep[e.g.][]{burgh07,sonnentrucker07}. 

Molecular gas occupies only a small volume fraction of the ISM, and %, and this appears to be no less true in supershell walls. 
probes of other physical regimes are needed to form a complete picture of a supershell system \citep[see e.g.][]{heiles99}. H{\sc i} observations in particular are indispensable in probing the structure and kinematics of the neutral ISM, and in correctly relating molecular clouds to the global structure of a supershell system. Much of the work discussed below combines H{\sc i} and CO data to investigate the structure and evolution of the ISM in supershells. Other tracers such as H$\alpha$ and soft X-rays are also useful, and provide valuable information on the ionised inner walls and the hot diffuse medium in shell interiors. 

A key morphological/kinematic indicator of potential cloud formation is co-moving CO clouds that form coherent parts of an atomic shell -- either lying along H{\sc i} walls or well embedded within them. %or are well embedded within, H{\sc i} walls, and form coherent parts of the atomic shell. 
The converse is molecular clouds that show a head-tail morphology or other signs of dynamical disruption, or are entrained in a shell interior, likely indicating pre-existing dense gas. % interacting with the shell. 
This simple diagnostic is by no means perfect. Various hybrid scenarios in which molecular clouds are formed early in a system's evolution and later impacted by subsequent supernova blast waves are certainly possible. Moreover, analysis can be frustrated by the irregularity of real observed supershells and line of sight confusion. Nevertheless, morphological diagnostics are an excellent starting point. %Moreover, in an inhomogeneous ISM populated by irregular shells and complicated (at least in the Milky Way) by line-of-sight confusion, its not always straightforward to assign a cloud to a given superstructure or to make a reliable morphological diagnosis.  %, or 
%reliably make this kind of morphological diagnosis.

%The converse being material in interior, moving slower, showing disrupted morphology, at end of tails etc. Though these by no means perfect -- hybrid scenarios such as molecules forming in RT drips then becoming disrupted by subsequent explosions are at least in theory possible. Plus in a frothy ISM populated by irregular, interacting shells, its not always straightforward to assign a cloud to a given shell, and/or to make this kind of diagnosis. This kind of morphological/kinematic discrimination more useful in regions that are predominantly atomic (for particularly molecule rich regions shell may expand into medium that is already mostly dense, cold. Then would be harder to distinguish between new and old material since there wouldn't be that large density contrast that results in pre-existing stuff becoming entrained, forming tails etc. This kind of consideration allows you to decide whether its likely that a given cloud or population of clouds were formed in-shell, though very hard to ever be sure for any given object. Additional information provided by total column densities providing constraints on shielding for molecules can help make a case for whether a cloud can survive or is being stripped/destroyed. More on that in section ..
A complementary approach is to measure whether supershells are associated with a net increase in the molecular fraction of the ISM in the volumes they occupy. 
This attempts to answer the question of whether the formation of new molecular gas has dominated over the destruction of pre-existing clouds, as measured at a particular epoch in the system's evolution. This is an approach taken by \citet[][discussed below]{dawson11a,dawson12} by comparing the H$_2$ mass fraction in supershell volumes with nearby control regions.

%This is a slightly different question to whether molecular cloud formation occurs at all -- it essentially asks whether formation dominates over destruction. However, a positive answer proves the former. This is an approach taken by Dawson et al. ... - comparing the molecular fraction in shell volumes with nearby `control' regions. 
Finally, stellar population studies are a powerful tool for reconstructing the history of a system. %, and can provide valuable constraints on the nature of any triggering that may have occurred. %\textbf{[Wind this up...]}%The combination of stellar age estimates, ... can in principle provide strong 
As well as providing valuable constraints on the original energy input from the central cluster, they can also provide insight into the nature and timing of any triggering that may have occurred during a supershell's evolution.  

% certainly in terms of the star formation history, which can itself provide strong clues to the form of triggering (if any) -- whole region sweep-up and simultaneous collapse vs local triggering of pre-existing gas interacting with shell shock.

%This section we're going to discuss various 

\subsection{The Local ISM}
\label{obslocal}

The local ISM, within a few hundred parsecs of the Solar System, is in many ways a microcosm of much of the Milky Way Disk. It is an environment that has been carved out and shaped by multiple generations of stellar activity -- characterised by a complex structure of loops, filaments, tunnels, irregular %interlocking 
shells and bubbles; and populated by the stellar clusters and associations variously responsible for and triggered by this activity \citep[see][]{degeus92,heiles98,lallement03,frisch11}. % various stellar clusters presumably responsible, with a complex structure, the details of which are being constantly refined and updated.
Unsurprisingly, the distribution of the local molecular ISM is closely related to these superstructures.%stellar feedback 
%structures. [Must be described in context of these....] 

Perhaps the most striking example is the Gould Belt and its H{\sc i} counterpart the Lindblad Ring. The Gould Belt/Lindblad Ring is the expanding, inclined ring of neutral gas and star-forming regions that contains all of the major local molecular cloud complexes within a distance of $\sim500$ pc from the Sun, including Orion, Taurus, Perseus, Ophiuchus and Lupus (\citealp{taylor87}; see also review by \citealp{poppel97}). %The Gould Belt has clearly dominated the history of the local ISM, and s
If the Gould Belt is a feedback structure, it is a superb candidate for the formation of molecular gas and stars from a swept-up supershell. Indeed, one scenario holds that the Gould Belt clouds are the remnants of an old fossil supershell that has fragmented sometime in the last 15-25 Myr to form gravitationally bound star-forming complexes (\citealp{olano82}; see also \citealp{bally01}). However, while stellar feedback is a strong contender, the origin of the Gould Belt is unclear, with other candidates including the impact of a high-velocity cloud \citep{comeron92,comeron94} and the braking of a supercloud entering the spiral arm (\citealp{olano01}; see also \citealp{grenier04} and references within). %, and it has been modelled successfully as both a cylindrical shock and superbubble system \citep{morino99,perrot03}.

As this uncertainty suggests, %surrounding the origins of the Gould Belt 
%underscores the fact that %disentangling threads of observational evidence is difficult, and 
reconstructing a region's history is often a challenging exercise in astronomical forensics. Nevertheless, there are numerous indications that stellar feedback has played an important role in the evolution of the local molecular ISM. %While it is beyond the scope of this paper to summarise in detail the vast body of work on these local clouds, w
We will explore a few illustrative examples in the paragraphs to follow.  

%kinematics of the well-known local molecular cloud complexes, which 
%which appear to form parts of an expanding ring whose H{\sc i} counterpart is 

%Many prominent local molecular clouds = Gould Belt. 
%Disentangling threads of observational evidence to reconstruct region's history is a challenging exercise in astronomical forensics.

The Lupus and Ophiuchus molecular cloud complexes form part of the neutral ISM delimiting the cavity of the Local Bubble -- the old supershell in which the Sun is currently located \citep[see][]{lallement03}. Their distances are $\sim155$ pc and $\sim120$ pc respectively \citep{lombardi08}, and both complexes have estimated (CO-bright) molecular masses %in each complex is estimated from CO line observations to be 
of $\sim10^4~M_{\odot}$ \citep{degeus91,tachihara01}, meaning that they are smaller than typical GMCs. %on the scale of . %, making them only moderately sized clouds. 
The Lupus clouds are located between the Upper-Centaurus-Lupus and Upper-Sco subgroups of the Sco-Cen OB association, apparently lying on the edge of an H{\sc i} shell blown by the latter (and younger) of these \citep{degeus92}. Parts of the cloud complex show evidence of interaction with the expanding shell, and it has been suggested that star formation may have been triggered by both the current interaction and a previous shock wave from the older Upper-Centaurus-Lupus shell %through the clouds 
several Myr ago \citep{tachihara01,merin08,tothill09}. An implication of this is that the molecular matter -- or at least some of it -- pre-dated these shells. Along similar lines, the detailed work on the Ophiuchus region by \citet{degeus92} stresses the interaction of the Upper-Sco subgroup with the much-studied $\rho$-Oph cloud. Here, molecular gas clearly pre-dates the current interaction, and is being stripped from the cloud to form streamers and tails. However, he also suggests the possibility of in-situ formation in the shell walls for some other parts of the complex. This is consistent with an alternative scenario suggested by \citet{preibisch07}, in which feedback flows formed both the present Lupus and Ophiuchus complexes, and the Lupus clouds were formed sandwiched at the interface where the two shells meet. This scenario is attractive in that it %explains the %star formation 
%history of the region %better than the picture of triggering in existing molecular clouds, which would 
%in a self-consistent way, and 
does not require molecular gas to exist for an unrealistically long time prior to the onset of star formation. %he interaction with shocks from the stellar groups. 
The authors also stress the role of multiple episodes of energy input, with clouds initially being formed during the wind bubble stage and then swept over to trigger star formation during a later supernova phase. It is interesting to note that this is also a solution proposed by \citet{inoue09} to overcome the difficulty of achieving high enough densities for molecular gas and star formation in a magnetically supported cloud. %Along similar lines, t

Another interesting Gould Belt complex is the Cepheus Flare. This region contains $\sim5\times10^3~M_{\odot}$ of molecular gas that forms part of an expanding shell of $R\sim50$ pc enclosing an old SNR \citep{grenier89,olano06}. \citet{kun98} note that the location of star forming sites at the edges of the CO clouds suggest external shock-triggering, which \citet{olano06} attribute to the same energetic event that created the shell -- again, tentatively suggesting that the clouds pre-date the current interaction. Interestingly, this shell may be associated with the prominent H{\sc i} ring known as the North Celestial Pole Loop \citep{meyerdierks91}, which itself contains the well-known molecular cirrus cloud, the Polaris Flare \citep{heithausen90}. The Polaris Flare is a classic example of a translucent molecular cloud -- diffuse \citep{meyerdierks96}, gravitationally unbound \citep{heithausen90} and non-star-forming \citep{andre10}. It has been described as `the archetype of the initial phases of molecular cloud formation' \citep{miville10}. In this context it is especially interesting to note that the Polaris Flare may be material swept up by the Cepheus Flare shell.

%At the top of the Cepheus Flare shell, between b = 15? and 40?, there is a prominent H I ring called the North Celestial Pole Loop (Heiles 1989; Meyerdierks, Heithausen & Reif 1991) or the Po- lar Ridge (Fejes & Wesselius 1973). This H I complex is related with a large molecular cirrus cloud known as the Polaris Flare (Heithausen & Thaddeus 1990; Heithausen et al. 1993; Meyerdierks & Heithausen 1996). This gas ridge can be interpreted as a product of the material ejected by a chimney effect from the Cepheus Flare shell.

%located at a distance of $\sim300$ pc with a molecular these clouds
%\begin{comment}
Somewhat further afield, \citet{heiles98} %report the discovery of a large nearby supershell GSH 238+00+09, located at a distance of $\sim800$ pc with semi-major and semi-minor axes of $\sim550$ and $\sim220$ pc in the plane of the Galaxy. He 
suggest a scenario in which star formation in the Orion and Gum regions (both at $D\sim500$ pc) was triggered by a more distant supershell GSH 238+00+09; 
%supershell GSH 238+00+09, at a distance of 800 pc, was responsible for triggering star formation in the nearby Orion and Gum regions, 
although it is unclear whether molecular clouds were formed by the action of the shell or merely compressed by it. This secondary star formation later blew the Orion-Eridanus \citep[e.g.][]{heiles99} and Gum Nebula \citep[e.g.][]{yamaguchin99} bubbles, both of which are themselves associated with CO-bright gas, including many cometary globules -- presumably remnants the parent clouds of the new young clusters. %now interacting with the young stellar sources 
\citep[see][and references within]{bally98,yamaguchin99}. %This kind of propagating and bubble formation is likely very common throughout the Milky Way. 

\begin{figure*}[t]
\begin{center}
\includegraphics[scale=1.0]{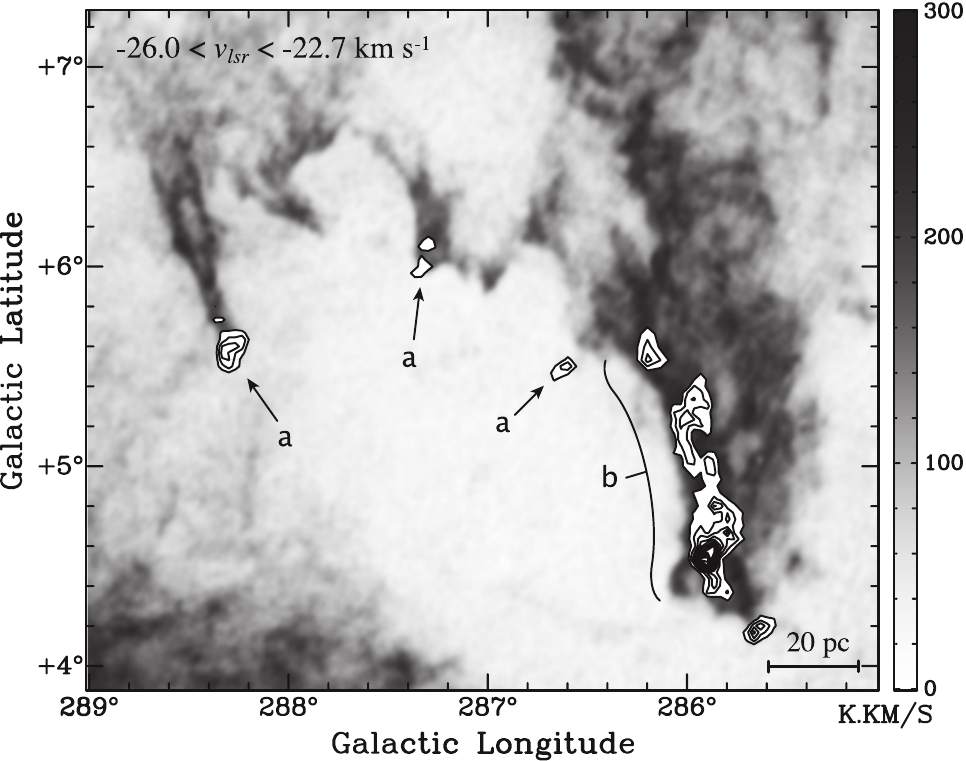}
\caption{\small Subsection of the wall of the Galactic supershell GSH 287+04--17 integrated over the velocity range indicated in the top-left corner of the image \citep[reproduced from][]{dawson12b}. The greyscale is H{\sc i} and the filled contours are $^{12}$CO(J=1--0). Features labelled a indicate small ($M_{\mathrm{H}_2}\sim10^{2\mathrm{-}3} M_{\odot}$) clumps of molecular gas offset towards the tips of atomic ÔfingersÕ that point in the direction of the shell centre. These are CO clouds that are likely being destroyed by their interaction with the shell. The feature labelled \textit{b} is an example of a larger ($M_{\mathrm{H}_2}\sim10^{4} M_{\odot}$) molecular cloud that is well embedded in atomic material, and forms a coherent part of the main shell wall. This is a strong candidate for in-situ formation of molecular gas in the shell wall. This entire region is located at $z\sim200$--300 pc above the Galactic midplane.}
\label{cfwall}
\end{center}
\end{figure*}

\subsection{The Milky Way Galaxy}
\label{obsmw}

In the wider Milky Way, a large number of studies have documented %the presence of 
molecular gas associated with %associated with %stellar-feedback superstructures 
supershells \citep{kundt87,koo88,heiles96,maciejewski96,normandeau96,jung96,bally98,patel98,rizzo98,yamaguchin99,fukui99,kim00,carpenter00,mcclure00,matsunaga01,moriguchi02,megeath03,moor03,dawson08a,dawson11a}, including notable examples of multiple generations of sequential star formation triggered by large-scale stellar feedback \citep{patel98,oey05}. However, as in the case of the local ISM, there is %almost always 
often ambiguity surrounding the origins of the molecular material itself. %Below we examine the few cases where the problem has been examined in detail.
Below we will discuss those objects for which the issue has been explicitly considered.

The Perseus Chimney is a vertically elongated ($\sim110\times230$ pc) supershell in the Perseus Arm of the Outer Galaxy, which is likely in the process of blowing out of the Galactic Disk \citep{normandeau96,dennison97,basu99}. The shell is bordered on its lower edge by the bright  W3/W4 H{\sc ii} regions, and is associated with substantial quantities of molecular gas, particularly around its base in the Galactic Plane \citep{digel96,heyer98}. \citet{heyer96} and \citet{taylor99} report on two cometary molecular clouds embedded in the chimney cavity, at altitudes of 20 pc and 50 pc and with molecular masses of $\sim5\times10^3$ and $\sim3\times10^3~M_{\odot}$. These clouds show strong evidence of interaction with the stellar winds and UV radiation of the IC 1805 cluster at the chimney base, including streamers and tails of both H{\sc i} and CO pointing directing away from the ionising sources. They are both well interpreted as remnants of an original molecular complex that was destroyed in the formation of the chimney. For the large \citep[$\sim10^5~M_{\odot}$;][]{digel96} reservoir of molecular gas around the base of the supershell, interpretations are less clear. Star formation along the edge of the shell in material bordering the W4 region implies compression and triggering, with the most natural explanation being that this occurs in pre-existing molecular gas \citep{carpenter00,oey05}. However, whether all of this material represents remnants of an older GMC complex, or whether additional episodes of compression have supplemented the molecular gas reservoir in the past is not clear. Some hint that this might have been the case comes from the suggestion of \citet{oey05} that the IC 1805 cluster itself may represent a later generation of star formation. In this picture its formation was triggered by an older progenitor cluster responsible for blowing an even older and larger supershell of which the current Perseus Chimney forms only the lower part \citep{reynolds01}. 

Several other studies have considered molecular cloud formation in a more quantitative way. \citet{jung96} report the association of a large mass of molecular gas ($\sim1.1\times10^6~M_{\odot}$) with the Outer Galaxy supershell GS 234--2, which they interpret in the context of the gravitational fragmentation of the swept-up atomic medium based on the analytical expression of \citet{mccray87}. %However, low spatial resolution means that a strong case cannot be made. 
\citet{kim00} make a more careful analysis of two small ($\sim10^3~M_{\odot}$) molecular clouds well-embedded within the H{\sc i} of the `Galactic Worm' GW 46.4+5.5, which makes up the vertical wall of a large ($340\times540$ pc) and relatively local ($D\sim1.4$ kpc) H{\sc i} shell. They also use the analytic expression for gravitational fragmentation, and find that the timescale for its onset is approximately equal to the estimated kinematic age of the shell ($t\sim5$ Myr), leading them to conclude that the molecular gas has likely formed from the swept-up ambient medium. In this context it is interesting that the molecular clouds (as measured from CO) do not appear to be globally gravitationally bound. While the molecular portion of the ISM presumably does not account for the entire mass of the `fragment' in which it is found, it is worth noting that unbound clouds are not inconsistent with the predictions of the colliding flows picture of molecular cloud formation. %\citet{patel98} also consider gravitational fragmentation of a swept-up shell in interpreting a large mass of molecular gas ($\sim1\times10^5~M_{\odot}$) associated with the Cepheus Bubble. However, in this case they interpret this material as surviving remnants of molecular gas that evolved from the original shell. 

\subsubsection{GSH 287+04--17 and GSH 277+00+36: Detailed Case Studies and Quantitative Analysis}
\label{mememe}

The most detailed analysis of the origin and evolution of molecular gas in Milky Way supershells is given in a series of papers by \citet{dawson08b,dawson08a,dawson11b,dawson11a}. These authors present matched resolution, parsec-scale observations of H{\sc i} and CO in two Galactic supershells, GSH 287+04--17 and GSH 277+00+36, enabling detailed investigation of the relationship between the atomic and molecular ISM in shell walls. They find substantial quantities of co-moving molecular gas in the H{\sc i} shells, with rich substructure in both tracers, including molecular gas seen elongated along the inner edges of the atomic walls, embedded within atomic filaments and clouds, or taking the form of small CO clouds at the tips of tapering atomic `fingers.' Figure \ref{cfwall} shows a section of the wall of GSH 287+04--17 that illustrates these features. Small CO clouds at the tips of H{\sc i} fingers have no substantial atomic envelopes and show no evidence for hidden `dark' material (either optically thick H{\sc i} or CO-dark H$_2$), implying that insufficient material exists to shield CO against photodissociation. Some also show shock-disrupted morphology, leading to the interpretation that these small clouds are pre-existing molecular gas currently undergoing dynamical disruption, gas stripping, and eventual dissociation due to their interaction with the shell. Survival lifetimes are roughly estimated as a few Myr, both from reference to the numerical results described in \S\ref{destruction} and through estimates of the photodissociation timescale. 

Conversely, CO clouds well embedded within the main atomic shell walls are excellent candidates for newly formed clouds. The feature labelled `b' in the figure is the strongest example of such an object. This cloud shows evidence for a substantial dark component identified from $100\mu$m IR excess, which comprises over 50\% of the total mass of the H{\sc i}-CO complex, and provides sufficient material to shield CO molecules against UV dissociation, demonstrating that the cloud can survive and continue to grow in its present form. The mean initial density required to sweep up the complex from the ambient medium is $\sim1$--10 cm$^{-3}$ (depending on the assumed geometry), consistent with a realistic mixture of WNM and CNM. Comparing with the numerical simulations of \S\ref{flows}, and assuming a flow speed equal to the present-day expansion velocity of $v_{exp}\sim10$ km s$^{-1}$, this implies that timescales of $<10^7$ yr are needed to form significant quantities of CO -- consistent with the estimated age of the shell. In addition, \citet{wunsch12} argue that the mass spectrum of molecular clumps in this region is consistent with the predictions of pressure-assisted gravitational instability in an expanding shell \citep[PAGI;][]{wunsch10}, providing further constraints on the physics of the swept-up ISM.  

The net effect of the two shells on the volumes they occupy is estimated by comparing the molecular fraction, $f_{\mathrm{H}_2} = M_{\mathrm{H}_2} / (M_{\mathrm{HI}} + M_{\mathrm{H}_2})$, in shell volumes -- including the evacuated voids -- to that in their local vicinities (essentially a proxy for the undisturbed medium). Since $f_{\mathrm{H}_2}$ varies with location in the Galactic Disk, these `background' regions are restricted carefully to include only emission that is genuinely local to the shells. The results of this analysis reveal that $f_{\mathrm{H}_2}$ in the supershell volumes is enhanced by a factor of $\sim2$--3 with respect to their local surroundings, implying that as much as $50$--$70\%$ of the molecular matter in their walls may have been formed as a direct result of stellar feedback. At present this analysis is restricted to two objects -- both of which were selected in part \textit{because} of their association with molecular clouds -- and no strong conclusions can be drawn about the Milky Way as a whole. Nevertheless, this is a compelling proof of concept, and some of the first quantitative evidence of molecular cloud formation in shell walls. 

\subsubsection{Cold H{\sc i} in Shell Walls}

The presence of cold atomic medium in supershell walls is important, since it is a necessary precursor to molecular gas. %Observational confirmation can be challenging, however, since the CNM is best observed in absorption, which requires the presence of a bright background source against which the cold foreground H{\sc i} can absorb -- either bright H{\sc i} emission or bright continuum. 
\citet{heiles82} report measurements of H{\sc i} absorption spectra in the walls of several nearby supershells, taken towards background continuum sources. They find that H{\sc i} in the shells is in the cold, neutral regime, with excitation temperatures ranging from 35--200 K -- much colder than temperatures outside the shell walls. 

\citet{dawson11a} fit multiple Gaussian profiles to H{\sc i} emission in selected subsections the walls of the two shells discussed in \S\ref{mememe}, and use the measured linewidths to put a strict upper limit on the kinetic temperature of the gas of $\sim350$ K. Assuming turbulence contributes approximately half of the observed linewidth \citep[e.g.][]{heiles03}, an estimate of $T_K\sim100$ K is found for the fitted sections of shell wall. Together with an estimated density of $\sim10$ cm$^{-3}$ this implies that the atomic shell walls are dominated by cold gas with parameters close to the canonical values for the CNM \citep[see also][]{mcclure03}. 

H{\sc i} self-absorption (HISA) provides a better probe of the morphology and physical properties of cold atomic clouds, although the requirement of bright background emission means that HISA is typically only observable at low Galactic latitudes. \citet{moss12} report the discovery of the very large supershell GSH 006--15+7, whose lower regions are seen as a striking cone-like absorption feature against the Galactic Plane. They estimate optical depths and kinetic temperatures of $\tau\sim3$ and $T_K\sim40$ K, respetively, again indicating the presence of copious quantities of cold, opaque atomic material.   %, limiting its usefulness in supershell work. of the Galactic Plane 
A similar and more extreme result is reported by \citet{knee01}, who present evidence for an extremely massive ($\sim10^7~M_{\odot}$), extremely cold ($\sim10$--20 K) atomic arc believed to belong to the distant outer galaxy shell GSH 139--03--69. 

Although little work on the thermal state of the atomic medium in supershells has been carried out beyond these studies, they provide strong evidence that much of the ISM in supershells is in the form of cold atomic gas, consistent with them playing an important role in ISM cooling and cloud formation. 

%Confirming cold HI in shell walls is important since it's the precursor to molecular gas. Hard to do since typically need absorption spectra to get at temperature of HI and separate cold from warm components, and for this you need bright background sources. Dawson and McClure-Griffiths confirm narrow spectra and put upper limit on T in selected sections of walls where spectra are unconfused. T < few 100 K. Nothing quantitative though in terms of cold mass fraction. Moss et al. have lovely HISA. And the dude with the cold HI shell nature/science paper.

\subsubsection{High-altitude Molecular Material}

Another role of supershells may be in supporting a molecular `thick disk' \citep{dame94,malhotra94b} by elevating molecular material well above its normal scale height \citep[$\sigma_z\sim60$ pc at the solar circle][]{malhotra94a}. Several studies have explicitly noted high-altitude molecular gas associated with feedback superstructures. \citet{megeath03} report the association of the large, high-altitude ($z\sim300$ pc) star-forming molecular cloud complex NGC281 with an expanding supershell. Both of the shells described in detail in \S\ref{mememe} also contain moderately sized ($\sim10^3$--$10^4~M_\odot$) molecular clouds at heights of up to $z\sim450$ pc above the midplane \citep{dawson11a}. This includes objects nominally identified both as pre-existing and newly formed clouds.

\citet{matsunaga01} also report the detection of eight supershell candidates initially identified from CO alone, mostly seen either as holes in the CO distribution or as strings of discrete molecular clouds that trace arc-like shapes above the plane.  %While lack of multiwavelength counterparts (particularly in H{\sc i})  confirmation from other wa, supershells provide a good explanation for most of the 
%While this sample is biassed towards high-altitude structures due to the fact that shell candidates are selected based on the presence of characteristic off-plane arcs, it is nevertheless interesting to note that 
Six of these objects contain molecular gas at $z\gtrsim150$ pc. Moderately elevated clouds are also seen by \citet{kim00} and \citet{moriguchi02} in GW 46.4+5.5 and the M16/M17 supershell, respectively. Finally, \citet{heiles96} note the presence of `high-z CO' associated with two Galactic Worms, GW 30.5--2.5 and GW 49.1--1.4, though they do not quote $z$ itself. These results suggest that the molecular thick disk is supported at least in part by both the formation and the displacement of molecular gas due to discrete large-scale stellar feedback events.  

%Megeath. Moriguchi? Me. Kim00, how about Patel? Matsunaga? Clearly important in providing molecular material off plane. Elevation of ISM by discrete feedback events a possible explanation for Dame and Thaddeus thick disk and possibly has interesting implications for Halo stars if material star forming. Dawson papers note star formation in clouds 250 pc off midplane. 

\subsubsection{Remarks on Molecular Fraction and Galactocentric Radius}

\begin{figure}[t]
\includegraphics[scale=0.3, angle=0]{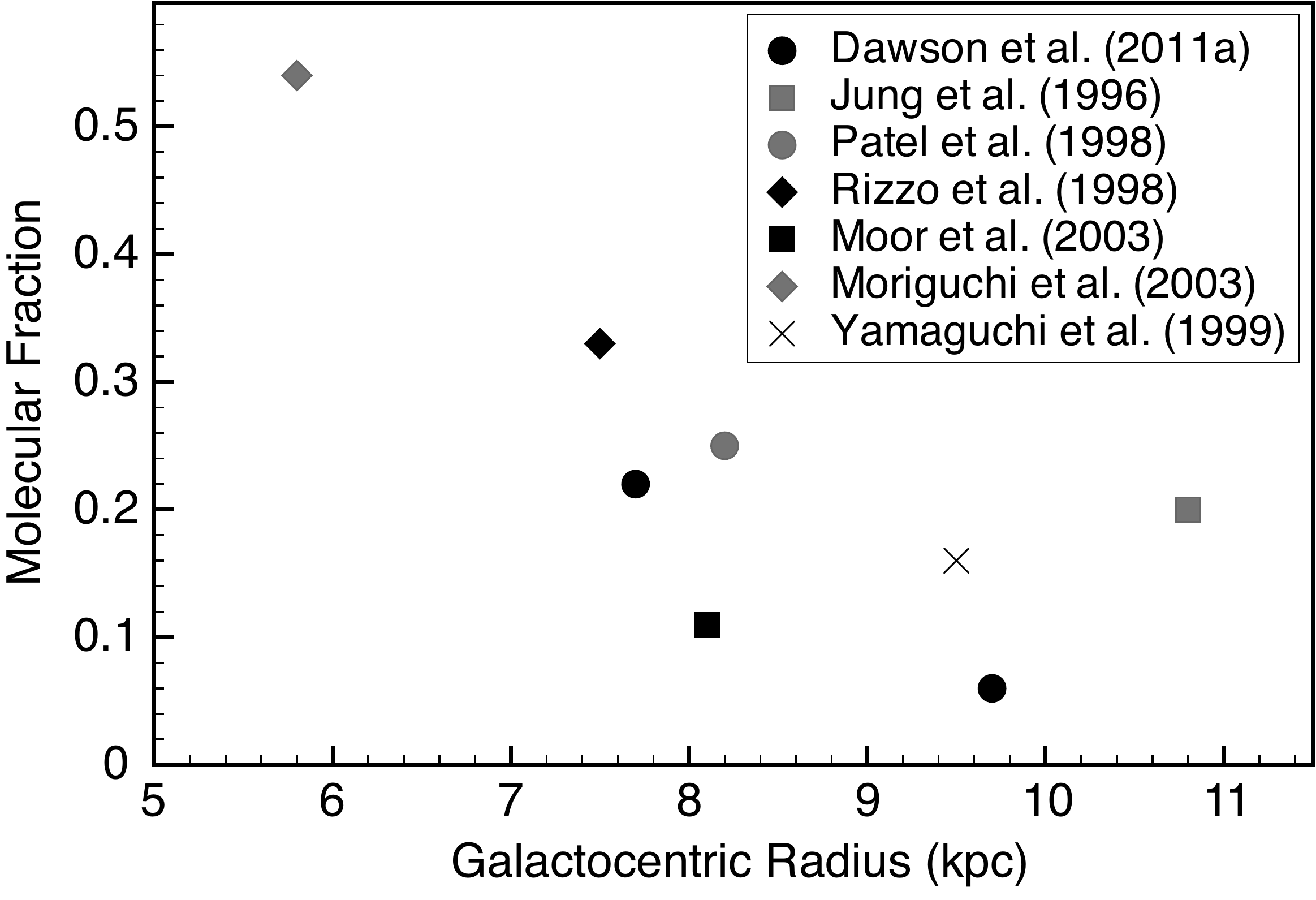}
\caption{\small The relationship between Galactocentric radius and molecular fraction in Milky Way supershells. The molecular fraction, $f_{\mathrm{H}_2} = M_{\mathrm{H}_2} / (M_{\mathrm{HI}} + M_{\mathrm{H}_2})$ is calculated for all supershells in the literature for which estimates of both atomic and molecular mass are available for the full shell.}
\label{fmolfig}
\end{figure}

\begin{figure*}[t]
\begin{center}
\includegraphics[scale=0.8, angle=0]{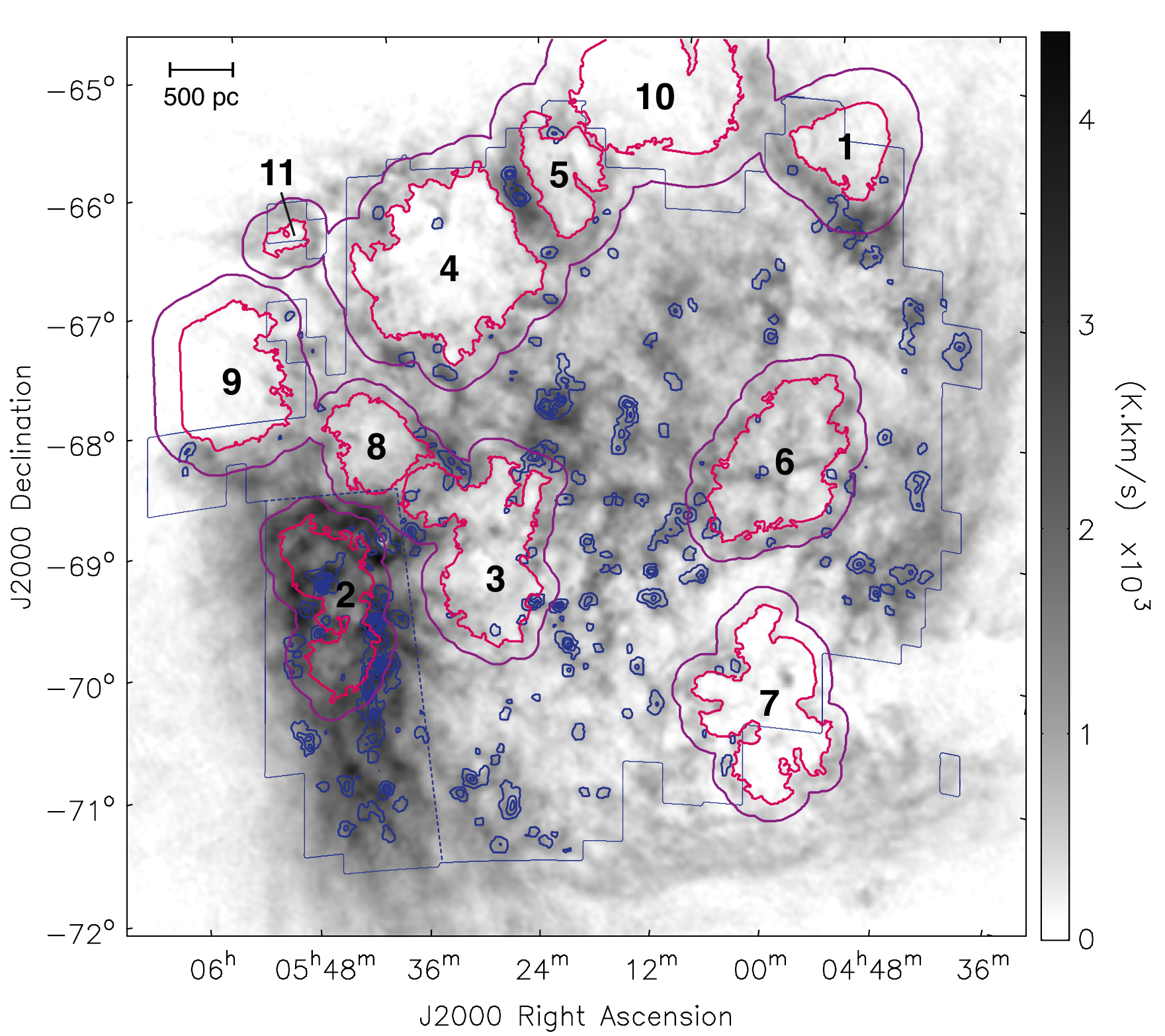}
\caption{\small Supergiant shells and shell complexes overlaid on integrated intensity map of the LMC \citep[reproduced by permission of the AAS from][]{dawson12}. The greyscale image is H{\sc i} \citep{kim03}, and blue contours are $^{12}$CO(J=1--0) \citep{fukui08}, processed as described in \citet{dawson12}, and integrated over the full velocity range of the LMC disk. The solid blue line marks the boundary of the region observed in CO. Dark pink lines trace the inner rims of the shell complexes and purple lines mark their outer boundaries (delineating the outer edges of the dense shells). Dotted blue lines enclose the region known as the southeastern H{\sc i} overdensity.}
\label{lmcfig}
\end{center}
\end{figure*}

The picture of supershell-driven molecular cloud formation outlined in this review applies primarily to a regime in which the ambient medium contains a large atomic component. Mean densities must also be high enough that feedback flows are able to accumulate sufficient quantities of material for gas cooling and molecule formation. %that is sufficiently dense for stellar feedback flows to accumulate the column densities required for molecular cloud formation. 
These conditions are applicable within a few kpc of the solar circle, but are unlikely to apply to the extreme inner or outer regions of the disk. The relevance of feedback-triggered molecular cloud formation is therefore likely to be a function of Galactocentric radius. %, and that there is a `sweet spot' at which it is most effective.

While quantitative measurements of molecular cloud formation in supershells are still too few to provide any observational insight into its variation with location, it is worth noting that available estimates of the H{\sc i} and H$_2$ masses of Galactic supershells do demonstrate the expected propensity for increasing molecular fraction with decreasing Galactocentric radius (see figure \ref{fmolfig}). This also has implications for the initial detection and statistical study of populations of supershells, particularly when detections are based on a single tracer. The eight supershell candidates discovered in CO by \citet{matsunaga01} are all inner Galaxy objects, located between $R\sim5.5$--7.5 kpc (rescaled for $R_0=8.0$ kpc). Conversely, H{\sc i}-based detections may well be biassed against %this population of 
inner Galaxy shells, as decreasing ambient atomic gas fraction and increasing confusion renders coherent H{\sc i} structures more difficult to detect.

\subsection{The Large Magellanic Cloud}
\label{obslmc}

%\subsubsection{The Magellanic Clouds}
%The morphology-based approach to feedback-triggered cloud formation -- which relies on first identifying supershell structures and then examining the associated molecular ISM -- requires that the neutral ISM is resolved to scales of $\sim10$ pc. Until recently this has not been feasible   
%\subsubsection{The Local Group}
The Large Magellanic Cloud (LMC) is the nearest star-forming galaxy to the Milky Way ($D\sim50$ kpc), and an excellent laboratory for studying the interaction between stars and the ISM. Unlike in the Milky Way, line of sight confusion is not a significant problem, and the reliable identification of supershells is considerably simplified, enabling a more statistical approach to the question of molecular cloud formation. A large number of shells and supershells have been identified \citep[e.g.][]{davies76,meaburn80,chu90,kim99}, %and the SMC \citep{meaburn80,staveley97,hatzidimitriou05}, 
in a variety of observational tracers, and several studies have explicitly considered their relationship to the molecular ISM.

%The connection between supershells and the molecular ISM has been primarily studied in the LMC. %, whose nearly face-on orientation \citep[$i\sim35^{\circ}$;][]{vandermarel01} makes it an attractive target.
\citet{yamaguchi01b} make a statistical study of the effects of optically identified supergiant shells (SGSs; defined as those whose radii are larger than the disk scale height) on the formation of stars and molecular clouds. %, based on $^{12}$CO(J=1--0) data at a spatial resolution of $\sim40$ pc. 
They find that both the number density and mass density of $^{12}$CO(J=1--0) molecular clouds is enhanced by a factor of 1.5--2 at the edges of SGSs, which they interpret as evidence of cloud formation. \citet{book09} examine a subset of this SGS population in detail, and argue that both the compression of pre-existing dense clouds and the formation of new molecular gas are likely occurring in the supergiant shells LMC 1, 4, 5, and 6. The interacting shells LMC 4 and LMC 5 contain particularly convincing examples of both processes: A large, dense ridge of molecular material compressed between them forms a striking example of the classic picture of efficient cloud formation at the interface of two shells \citep[e.g.][]{hartmann01,ntormousi11}, while a smaller cometary CO cloud with an H{\sc i} tail exhibits the classic morphological signatures of the interaction of a shock front with pre-existing dense gas \citep[see also][]{yamaguchi01a}. % heading toward the center of the SGS.
H{\sc i} absorption spectra towards background continuum sources also suggest that efficient cooling has driven the production of large quantities of cold atomic gas in the walls of LMC 4 \citep{marx00}. This is consistent with the behaviour seen in Milky Way supershells.

\citet{dawson12} combine CO data with H{\sc i} synthesis images \citep{kim03} to make the first quantitative measurement of feedback-driven cloud formation in an entire galactic system. %explore this issue in more depth. They 
As described in \S\ref{mememe} for Milky Way shells, their method compares the molecular fraction, $f_{\mathrm{H}_2} = M_{\mathrm{H}_2} / (M_{\mathrm{HI}} + M_{\mathrm{H}_2})$, in SGS volumes with that elsewhere in the disk, in order to assess whether the passage of an SGS through the ISM is associated with a net increase in the production of molecular gas. Figure \ref{lmcfig} shows an H{\sc i} and CO map of the LMC disk, overlaid with the outlines of the SGS complexes used in this analysis. %, defined by the thresholding the inner rims of the atomic shell walls. % than would otherwise have been created. % if no SGS had been present. %he underlying assumption is that if the net effect of a supergiant shell is to produce more molecular matter in its swept-up walls than is destroyed by its initial passage through the ISM, then the volumes now occupied by SGSs should be more molecular than they were prior to the occurrence of the shell.
% in order to assess whether the passage of an SGS through the ISM causes the enhanced production of H$_2$ in the volumes it affects. 
They find that $f_{\mathrm{H}_2}$ in the aggregate volume occupied by all SGSs is identical to the rest of the LMC disk, suggesting that supergiant shells are not a dominant driver of molecular cloud formation on galactic scales. Indeed, the global structure of the LMC disk is found to be a better determinant of where the highest molecular fractions are found. However, the majority of objects ($\sim70\%$ by mass) are more molecular than their local surroundings, implying that the presence of a supergiant shell does on average have a small positive effect on the molecular gas fraction. This analysis is used to place a lower limit on the total fraction of molecular cloud formation in the LMC driven by large-scale stellar feedback, which is estimated to be $\gtrsim4$--$11\%$ of the total molecular mass of the galaxy. %Although very much a rough estimate, this is the first quantitative measurement of the contribution of feedback-driven cloud formation in a galactic system. 

%One of the most promising sites of new cloud formation is examined in detail by \citet{yamaguchi01a}.

The importance of stellar feedback as a driver of molecular cloud formation is expected to be dependent on galaxy type. As a dwarf irregular, the Large Magellanic Cloud (LMC) is expected to be more susceptible to feedback-triggered molecular cloud formation than early-type spirals such as the Milky Way \citep[][see also \S\ref{gravity}]{elmegreen02a}. With low shear \citep{weidner10}, a large H{\sc i} scale height \citep{brinks02} and weak spiral structure, shells in the LMC are able to grow larger before they are deformed, expand further before vertical blowout and depressurisation, and not suffer the spiral arm disruption that will affect their Milky Way counterparts. %The visible impact of stellar feedback on ISM structure is generally larger in dwarf galaxies than in grand design spirals. In the case of the LMC, t
%The low shear allows shells to grow larger before they are deformed, high scale heights allow them to expand further before vertical blowout and depressurisation, and the lack of a strong spiral potential mitigates against disruption by spiral arms. 
Similarly, the lack of strong spiral structure, combined with a relatively weak disk gravitational potential, suggests that galaxy-scale self gravity and the accumulation of the ISM in spiral arms may play a less important role in dense gas formation than in grand-design spirals.  %systems. %gravitational reduce the ability of large-scale disk perturbations and spiral arms to %one potential method of 
%accumulate material into dense clouds, suggesting that the contribution of stellar feedback to cloud formation may be greater than in grand-design spirals such as the Milky Way. \textbf{Edit me do.}

\section{Summary, Discussion and Future Directions}
\label{summary}

%What do we currently know about molecular cloud formation by large-scale stellar feedback? T
%There is a large body of literature documenting the many and varied interactions of stellar energy input with the ISM. 
The accumulation, compression and cooling of the ISM in supershells formed by OB cluster feedback can drive the production of star-forming molecular clouds.
Analytical and numerical models provide a strong theoretical basis for %the formation of molecular gas in expanding shells and feedback flows
this process, and observations leave little doubt that molecular cloud formation via large-scale feedback is occurring in both the Milky Way and the LMC. 

Theoretically, many details remain to be hammered out, including the ability of magnetic fields to prevent compressed material from reaching sufficient densities for molecule formation, whether multiple episodes of compression are necessary, and whether the initial ambient density must be higher than canonical WNM values in order for flow-driven cloud formation to occur. %in order for cloud formation to proceed on reasonable timescales 
\citep[e.g.][]{dobbs12,inoue12}. 
Observationally, there are many convincing candidates for molecular clouds formed in supershell walls, and ample evidence that the atomic ISM in these shell walls tends to be cold, demonstrating that the cooling of swept-up material occurs readily. Enhanced molecular gas fractions in the volumes occupied by two Galactic supershells have also provided one of the first quantitative measurements of feedback-driven cloud formation in the Milky Way \citep{dawson11b}. However, the question of how readily the process occurs remains open, as does its relative importance in the galactic context. %, remains open. 
The first direct measurement of the contribution of stellar feedback to molecular cloud formation on galactic scales has now been made in the LMC, where $\gtrsim4$--11\% of the total molecular gas content of the galaxy is estimated to have been formed as a direct result of large-scale feedback \citep{dawson12}. %Such measurements in the Milky Way have been limited to a small number of individual objects, h
However, no such estimates exist for the Milky Way or other nearby galaxies.

%Feedback flows also provide an attractive means of driving episodes of near-simultaneous star formation that are correlated over large spatial scales, thus explaining the observed ages and configurations of stellar populations in the solar neighbourhood \citep{hartmann01}. 

%Overall, there is little doubt that molecular cloud formation via large-scale feedback is occurring, both in the Milky Way, and in the LMC, where spatially resolved observations of CO clouds in H{\sc i} supershells provide convincing evidence of triggered cloud formation. However, the question of how readily the process occurs remains open, as does its relative importance in the galactic context. %, remains open. 
%One of the few direct measurements of the contribution of stellar feedback to molecular cloud formation on galactic scales has been made in the LMC, where $\gtrsim4$--11\% of the total molecular gas content of the galaxy is estimated to have been formed as a direct result of large-scale feedback \citep{dawson12}. Such measurements in the Milky Way have been limited to a small number of individual objects, however, and no such estimates exist for other nearby galaxies.

%Feedback flows certainly provide an attractive means of driving episodes of near-simultaneous star formation that are correlated over large spatial scales, and thus can explain the observed ages and configurations of stellar populations in the solar neighbourhood \citep{hartmann01}. 
Thinking globally, large-scale gravitational instabilities play a major role in the initial formation of dense gas in galaxy disks \citep[see e.g.][and references within]{elmegreen02b,maclow12}, and may be the primary driver of molecular gas formation in most systems. Spiral arms are also clearly an important means of accumulating material, although whether they drive a significant amount of molecular gas formation, or whether their primary influence is to agglomerate existing GMCs into larger complexes is not yet clear \citep{koda09}. A picture is suggested in which large-scale feedback plays a secondary role -- initiating further episodes of compression and additional cloud formation in an inhomogeneous medium that already contains significant amounts of dense structure. This picture is %consistent with the current picture, and is %\textbf{[Something about different galaxy types and the work of Stas's friend]} 
%also 
tentatively consistent with the results presented in this review. %However, stronger observational evidence is needed before this conclusion can be argued strongly.  
However, direct measurements of the contribution of stellar feedback to molecular cloud formation rates are  strongly desirable. %, however.  

Observational work that explicitly deals with feedback-triggered cloud formation has generally taken a morphology-based approach, which relies on first identifying supershell structures and then examining the associated molecular ISM. This strategy requires that the neutral ISM be resolved to scales of $\sim10$--100 pc, which has historically been challenging for external galaxies, particularly in molecular line observations. With the advent of millimetre and sub-millimetre interferometers such as the Combined Array for Research in Millimeter Astronomy (CARMA) and the Atacama Large Millimeter/submillimeter Array (ALMA), high-resolution, high-sensitivity observations of molecular tracers are now becoming possible in local group galaxies. This opens up the exciting possibility of extending the study of feedback-triggered molecular cloud formation to large samples of galaxies with a range of different properties. Such work will be essential in constructing a consistent model of the impact of stellar feedback on ISM evolution across a broad range of galaxy types. %in the evolution of galactic systems, and in providing insight into the braoder astrophysical impact of 

Gaseous disks contain large amounts of stochastic structure, %that can mimic feedback holes \citep{wada00}, 
and the reliable identification of signatures of stellar feedback -- particularly in the neutral ISM alone -- is a source of uncertainty in observational studies. %An important challenge for the future will be developing a 
A preferable strategy would be a means of exploring feedback-triggered cloud formation that does not first rely on the morphological identification of shell-like structures. 
%This is an area in which observers and theorists should work closely to devise statistical measures that can be extracted from observational data in order to constrain the origin of 
A promising approach is suggested by the work of \citet{dobbs12}, who identify characteristic distributions in the eigenvalues of the local rate-of-strain tensor in model disks dominated by different astrophysical drivers of converging flows. %: spiral arms, disk gravity or stellar feedback. %different astrophysical processes 
This suggests the possibility of identifying the primary drivers of the flows that assemble dense clouds %in observational data, which could in theory be used to discriminate between dominant cloud formation mechanisms 
without recourse to morphological arguments, and hence quantifying the importance of stellar feedback in a robust and statistical sense. While it remains to be seen if this method can be usefully applied to real astronomical datasets (which cannot directly provide information on the 3D velocity field of the ISM), such direct marriages between simulations and observations are likely to provide powerful tools for the interpretation of future high-resolution data. This underscores the importance of using model galaxies -- in which the input parameters are well constrained -- to develop observational diagnostics for the ISM in real systems.

%Can we also seek means of constraining cloud formation mechanisms that do not first rely on the morphological identification of shell-like structures? as numerical models make clear, it's not as simple as all that - we can have things that mimic shells and real feedback sites may not be obvious. A statistical measure would be great, e.g. Clare's flow parameters. Though these probably not that effective in 2D+v space of real observations. 

\section*{Acknowledgments} %If needed

My thanks to Eva Ntormousi and Alex Hill for their kind permission to make use of their figures in this paper.

\bibliographystyle{apj}
\bibliography{pasabib}

\begin{thebibliography}{199}
\expandafter\ifx\csname natexlab\endcsname\relax\def\natexlab#1{#1}\fi

\bibitem[{{Andr{\'e}} {et~al.}(2010){Andr{\'e}}, {Men'shchikov}, {Bontemps},
  {K{\"o}nyves}, {Motte}, {Schneider}, {Didelon}, {Minier}, {Saraceno},
  {Ward-Thompson}, {di Francesco}, {White}, {Molinari}, {Testi}, {Abergel},
  {Griffin}, {Henning}, {Royer}, {Mer{\'{\i}}n}, {Vavrek}, {Attard},
  {Arzoumanian}, {Wilson}, {Ade}, {Aussel}, {Baluteau}, {Benedettini},
  {Bernard}, {Blommaert}, {Cambr{\'e}sy}, {Cox}, {di Giorgio}, {Hargrave},
  {Hennemann}, {Huang}, {Kirk}, {Krause}, {Launhardt}, {Leeks}, {Le Pennec},
  {Li}, {Martin}, {Maury}, {Olofsson}, {Omont}, {Peretto}, {Pezzuto}, {Prusti},
  {Roussel}, {Russeil}, {Sauvage}, {Sibthorpe}, {Sicilia-Aguilar}, {Spinoglio},
  {Waelkens}, {Woodcraft}, \& {Zavagno}}]{andre10}
{Andr{\'e}}, P., {et~al.} 2010, \aap, 518, L102

\bibitem[{{Audit} \& {Hennebelle}(2005)}]{audit05}
{Audit}, E., \& {Hennebelle}, P. 2005, \aap, 433, 1

\bibitem[{{Bagetakos} {et~al.}(2011){Bagetakos}, {Brinks}, {Walter}, {de Blok},
  {Usero}, {Leroy}, {Rich}, \& {Kennicutt}}]{bagetakos11}
{Bagetakos}, I., {Brinks}, E., {Walter}, F., {de Blok}, W.~J.~G., {Usero}, A.,
  {Leroy}, A.~K., {Rich}, J.~W., \& {Kennicutt}, Jr., R.~C. 2011, \aj, 141, 23

\bibitem[{{Bally}(2001)}]{bally01}
{Bally}, J. 2001, in Astronomical Society of the Pacific Conference Series,
  Vol. 231, Tetons 4: Galactic Structure, Stars and the Interstellar Medium,
  ed. C.~E. {Woodward}, M.~D. {Bicay}, \& J.~M. {Shull}, 204

\bibitem[{{Bally} {et~al.}(1998){Bally}, {Theil}, \& {Sutherland}}]{bally98}
{Bally}, J., {Theil}, D., \& {Sutherland}, R., S. 1998, in {The Orion Complex
  Revisited}, ed. M.~{McCaughrean} \& A.~{Burkert}, Astronomical Society of the
  Pacific Conference Series

\bibitem[{{Banerjee} {et~al.}(2009){Banerjee}, {V{\'a}zquez-Semadeni},
  {Hennebelle}, \& {Klessen}}]{banerjee09}
{Banerjee}, R., {V{\'a}zquez-Semadeni}, E., {Hennebelle}, P., \& {Klessen},
  R.~S. 2009, \mnras, 398, 1082

\bibitem[{{Basu} {et~al.}(1999){Basu}, {Johnstone}, \& {Martin}}]{basu99}
{Basu}, S., {Johnstone}, D., \& {Martin}, P.~G. 1999, \apj, 516, 843

\bibitem[{{Bergin} {et~al.}(2004){Bergin}, {Hartmann}, {Raymond}, \&
  {Ballesteros-Paredes}}]{bergin04}
{Bergin}, E.~A., {Hartmann}, L.~W., {Raymond}, J.~C., \& {Ballesteros-Paredes},
  J. 2004, \apj, 612, 921

\bibitem[{{Book} {et~al.}(2009){Book}, {Chu}, {Gruendl}, \& {Fukui}}]{book09}
{Book}, L.~G., {Chu}, Y.-H., {Gruendl}, R.~A., \& {Fukui}, Y. 2009, \aj, 137,
  3599

\bibitem[{{Boss}(1995)}]{boss95}
{Boss}, A.~P. 1995, \apj, 439, 224

\bibitem[{{Bournaud} {et~al.}(2010){Bournaud}, {Elmegreen}, {Teyssier},
  {Block}, \& {Puerari}}]{bournaud10}
{Bournaud}, F., {Elmegreen}, B.~G., {Teyssier}, R., {Block}, D.~L., \&
  {Puerari}, I. 2010, \mnras, 409, 1088

\bibitem[{{Brand} \& {Zealey}(1975)}]{brand75}
{Brand}, P.~W.~J.~L., \& {Zealey}, W.~J. 1975, \aap, 38, 363

\bibitem[{{Brinks} {et~al.}(2002){Brinks}, {Walter}, \& {Ott}}]{brinks02}
{Brinks}, E., {Walter}, F., \& {Ott}, J. 2002, in Astronomical Society of the
  Pacific Conference Series, Vol. 275, Disks of Galaxies: Kinematics, Dynamics
  and Peturbations, ed. E.~{Athanassoula}, A.~{Bosma}, \& R.~{Mujica}, 57--60

\bibitem[{{Bruhweiler} {et~al.}(1980){Bruhweiler}, {Gull}, {Kafatos}, \&
  {Sofia}}]{bruhweiler80}
{Bruhweiler}, F.~C., {Gull}, T.~R., {Kafatos}, M., \& {Sofia}, S. 1980, \apjl,
  238, L27

\bibitem[{{Carpenter} {et~al.}(2000){Carpenter}, {Heyer}, \&
  {Snell}}]{carpenter00}
{Carpenter}, J.~M., {Heyer}, M.~H., \& {Snell}, R.~L. 2000, \apjs, 130, 381

\bibitem[{{Chu} \& {Mac Low}(1990)}]{chu90}
{Chu}, Y.-H., \& {Mac Low}, M.-M. 1990, \apj, 365, 510

\bibitem[{{Clark} {et~al.}(2012){Clark}, {Glover}, {Klessen}, \&
  {Bonnell}}]{clark12}
{Clark}, P.~C., {Glover}, S.~C.~O., {Klessen}, R.~S., \& {Bonnell}, I.~A. 2012,
  \mnras, 424, 2599

\bibitem[{{Comeron} \& {Torra}(1992)}]{comeron92}
{Comeron}, F., \& {Torra}, J. 1992, \aap, 261, 94

\bibitem[{{Comeron} \& {Torra}(1994)}]{comeron94}
---. 1994, \aap, 281, 35

\bibitem[{{Cowie} \& {McKee}(1977)}]{cowie77}
{Cowie}, L.~L., \& {McKee}, C.~F. 1977, \apj, 211, 135

\bibitem[{{Dale} {et~al.}(2007){Dale}, {Bonnell}, \& {Whitworth}}]{dale07}
{Dale}, J.~E., {Bonnell}, I.~A., \& {Whitworth}, A.~P. 2007, \mnras, 375, 1291

\bibitem[{{Dale} {et~al.}(2012){Dale}, {Ercolano}, \& {Bonnell}}]{dale12}
{Dale}, J.~E., {Ercolano}, B., \& {Bonnell}, I.~A. 2012, \mnras, 427, 2852

\bibitem[{{Dale} {et~al.}(2009){Dale}, {W{\"u}nsch}, {Whitworth}, \& {Palou{\v
  s}}}]{dale09}
{Dale}, J.~E., {W{\"u}nsch}, R., {Whitworth}, A., \& {Palou{\v s}}, J. 2009,
  \mnras, 398, 1537

\bibitem[{{Dame} \& {Thaddeus}(1994)}]{dame94}
{Dame}, T.~M., \& {Thaddeus}, P. 1994, \apjl, 436, L173

\bibitem[{{Davies} {et~al.}(1976){Davies}, {Elliott}, \& {Meaburn}}]{davies76}
{Davies}, R.~D., {Elliott}, K.~H., \& {Meaburn}, J. 1976, \memras, 81, 89

\bibitem[{{Dawson} {et~al.}(2008{\natexlab{a}}){Dawson}, {Kawamura}, {Mizuno},
  {Onishi}, \& {Fukui}}]{dawson08b}
{Dawson}, J.~R., {Kawamura}, A., {Mizuno}, N., {Onishi}, T., \& {Fukui}, Y.
  2008{\natexlab{a}}, \pasj, 60, 1297

\bibitem[{{Dawson} {et~al.}(2011{\natexlab{a}}){Dawson}, {McClure-Griffiths},
  {Dickey}, \& {Fukui}}]{dawson11b}
{Dawson}, J.~R., {McClure-Griffiths}, N.~M., {Dickey}, J.~M., \& {Fukui}, Y.
  2011{\natexlab{a}}, \apj, 741, 85

\bibitem[{{Dawson} {et~al.}(2012){Dawson}, {McClure-Griffiths}, \&
  {Fukui}}]{dawson12b}
{Dawson}, J.~R., {McClure-Griffiths}, N.~M., \& {Fukui}, Y. 2012, in EAS
  Publications Series, Vol.~56, EAS Publications Series, ed. M.~A. {de
  Avillez}, 155--158

\bibitem[{{Dawson} {et~al.}(2011{\natexlab{b}}){Dawson}, {McClure-Griffiths},
  {Kawamura}, \& {Mizuno, N. et al.}}]{dawson11a}
{Dawson}, J.~R., {McClure-Griffiths}, N.~M., {Kawamura}, A., \& {Mizuno, N. et
  al.} 2011{\natexlab{b}}, \apj, 728, 127

\bibitem[{{Dawson} {et~al.}(2013){Dawson}, {McClure-Griffiths}, {Wong},
  {Dickey}, {Hughes}, {Fukui}, \& {Kawamura}}]{dawson12}
{Dawson}, J.~R., {McClure-Griffiths}, N.~M., {Wong}, T., {Dickey}, J.~M.,
  {Hughes}, A., {Fukui}, Y., \& {Kawamura}, A. 2013, \apj, 763, 56

\bibitem[{{Dawson} {et~al.}(2008{\natexlab{b}}){Dawson}, {Mizuno}, {Onishi},
  {McClure-Griffiths}, \& {Fukui}}]{dawson08a}
{Dawson}, J.~R., {Mizuno}, N., {Onishi}, T., {McClure-Griffiths}, N.~M., \&
  {Fukui}, Y. 2008{\natexlab{b}}, \mnras, 387, 31

\bibitem[{{de Avillez} \& {Berry}(2001)}]{avillez01}
{de Avillez}, M.~A., \& {Berry}, D.~L. 2001, \mnras, 328, 708

\bibitem[{{de Avillez} \& {Breitschwerdt}(2005)}]{avillez05}
{de Avillez}, M.~A., \& {Breitschwerdt}, D. 2005, \aap, 436, 585

\bibitem[{{de Geus}(1992)}]{degeus92}
{de Geus}, E.~J. 1992, \aap, 262, 258

\bibitem[{{de Geus} \& {Burton}(1991)}]{degeus91}
{de Geus}, E.~J., \& {Burton}, W.~B. 1991, \aap, 246, 559

\bibitem[{{Deharveng} {et~al.}(2010){Deharveng}, {Schuller}, {Anderson},
  {Zavagno}, {Wyrowski}, {Menten}, {Bronfman}, {Testi}, {Walmsley}, \&
  {Wienen}}]{deharveng10}
{Deharveng}, L., {et~al.} 2010, \aap, 523, A6

\bibitem[{{Dennison} {et~al.}(1997){Dennison}, {Topasna}, \&
  {Simonetti}}]{dennison97}
{Dennison}, B., {Topasna}, G.~A., \& {Simonetti}, J.~H. 1997, \apjl, 474, L31

\bibitem[{{Dickey}(2012)}]{dickey09}
{Dickey}, J.~M. 2012, in EAS Publications Series, Vol.~56, EAS Publications
  Series, ed. M.~A. {de Avillez}, 371--379

\bibitem[{{Digel} {et~al.}(1996){Digel}, {Lyder}, {Philbrick}, {Puche}, \&
  {Thaddeus}}]{digel96}
{Digel}, S.~W., {Lyder}, D.~A., {Philbrick}, A.~J., {Puche}, D., \& {Thaddeus},
  P. 1996, \apj, 458, 561

\bibitem[{{Dobbs} \& {Bonnell}(2008)}]{dobbs08}
{Dobbs}, C.~L., \& {Bonnell}, I.~A. 2008, \mnras, 385, 1893

\bibitem[{{Dobbs} {et~al.}(2006){Dobbs}, {Bonnell}, \& {Pringle}}]{dobbs06}
{Dobbs}, C.~L., {Bonnell}, I.~A., \& {Pringle}, J.~E. 2006, \mnras, 371, 1663

\bibitem[{{Dobbs} {et~al.}(2011){Dobbs}, {Burkert}, \& {Pringle}}]{dobbs11}
{Dobbs}, C.~L., {Burkert}, A., \& {Pringle}, J.~E. 2011, \mnras, 417, 1318

\bibitem[{{Dobbs} {et~al.}(2012){Dobbs}, {Pringle}, \& {Burkert}}]{dobbs12}
{Dobbs}, C.~L., {Pringle}, J.~E., \& {Burkert}, A. 2012, \mnras, 425, 2157

\bibitem[{{Draine} \& {Bertoldi}(1996)}]{draine96}
{Draine}, B.~T., \& {Bertoldi}, F. 1996, \apj, 468, 269

\bibitem[{{Efremov} {et~al.}(1999){Efremov}, {Ehlerov{\'a}}, \& {Palou{\v s}
  }}]{efremov99}
{Efremov}, Y.~N., {Ehlerov{\'a}}, S., \& {Palou{\v s} }, J. 1999, \aap, 350,
  457

\bibitem[{{Ehlerova} \& {Palous}(1996)}]{ehlerova96}
{Ehlerova}, S., \& {Palous}, J. 1996, \aap, 313, 478

\bibitem[{{Ehlerova} {et~al.}(1997){Ehlerova}, {Palous}, {Theis}, \&
  {Hensler}}]{ehlerova97}
{Ehlerova}, S., {Palous}, J., {Theis}, C., \& {Hensler}, G. 1997, \aap, 328,
  121

\bibitem[{{Ehlerov{\'a}} \& {Palou{\v s}}(2002)}]{ehlerova02}
{Ehlerov{\'a}}, S., \& {Palou{\v s}}, J. 2002, \mnras, 330, 1022

\bibitem[{{Ehlerov{\'a}} \& {Palou{\v s}}(2005)}]{ehlerova05}
---. 2005, \aap, 437, 101

\bibitem[{{Elmegreen}(2000)}]{elmegreen00}
{Elmegreen}, B.~G. 2000, \apj, 530, 277

\bibitem[{{Elmegreen}(2002)}]{elmegreen02b}
---. 2002, \apj, 577, 206

\bibitem[{{Elmegreen} {et~al.}(2002){Elmegreen}, {Palou{\v s}}, \&
  {Ehlerov{\'a}}}]{elmegreen02a}
{Elmegreen}, B.~G., {Palou{\v s}}, J., \& {Ehlerov{\'a}}, S. 2002, \mnras, 334,
  693

\bibitem[{{Federrath} \& {Klessen}(2012)}]{federrath12}
{Federrath}, C., \& {Klessen}, R.~S. 2012, \apj, 761, 156

\bibitem[{{Federrath} {et~al.}(2010){Federrath}, {Roman-Duval}, {Klessen},
  {Schmidt}, \& {Mac Low}}]{federrath10}
{Federrath}, C., {Roman-Duval}, J., {Klessen}, R.~S., {Schmidt}, W., \& {Mac
  Low}, M.-M. 2010, \aap, 512, A81

\bibitem[{{Ferri{\`e}re}(2001)}]{ferriere01}
{Ferri{\`e}re}, K.~M. 2001, Reviews of Modern Physics, 73, 1031

\bibitem[{{Field} {et~al.}(1969){Field}, {Goldsmith}, \& {Habing}}]{field69}
{Field}, G.~B., {Goldsmith}, D.~W., \& {Habing}, H.~J. 1969, \apjl, 155, L149+

\bibitem[{{Fragile} {et~al.}(2004){Fragile}, {Murray}, {Anninos}, \& {van
  Breugel}}]{fragile04}
{Fragile}, P.~C., {Murray}, S.~D., {Anninos}, P., \& {van Breugel}, W. 2004,
  \apj, 604, 74

\bibitem[{{Franco} \& {Cox}(1986)}]{franco86}
{Franco}, J., \& {Cox}, D.~P. 1986, \pasp, 98, 1076

\bibitem[{{Frisch} {et~al.}(2011){Frisch}, {Redfield}, \& {Slavin}}]{frisch11}
{Frisch}, P.~C., {Redfield}, S., \& {Slavin}, J.~D. 2011, \araa, 49, 237

\bibitem[{{Fukui} {et~al.}(1999){Fukui}, {Onishi}, {Abe}, {Kawamura},
  {Tachihara}, {Yamaguchi}, {Mizuno}, \& {Ogawa}}]{fukui99}
{Fukui}, Y., {Onishi}, T., {Abe}, R., {Kawamura}, A., {Tachihara}, K.,
  {Yamaguchi}, R., {Mizuno}, A., \& {Ogawa}, H. 1999, \pasj, 51, 751

\bibitem[{{Fukui} {et~al.}(2008){Fukui}, {Kawamura}, {Minamidani}, {Mizuno},
  {Kanai}, {Mizuno}, {Onishi}, {Yonekura}, {Mizuno}, {Ogawa}, \&
  {Rubio}}]{fukui08}
{Fukui}, Y., {et~al.} 2008, \apjs, 178, 56

\bibitem[{{Glover} \& {Clark}(2012)}]{glover12}
{Glover}, S.~C.~O., \& {Clark}, P.~C. 2012, \mnras, 426, 377

\bibitem[{{Glover} \& {Mac Low}(2007)}]{glover07}
{Glover}, S.~C.~O., \& {Mac Low}, M.-M. 2007, \apj, 659, 1317

\bibitem[{{Glover} \& {Mac Low}(2011)}]{glover11}
---. 2011, \mnras, 412, 337

\bibitem[{{Gregori} {et~al.}(1999){Gregori}, {Miniati}, {Ryu}, \&
  {Jones}}]{gregori99}
{Gregori}, G., {Miniati}, F., {Ryu}, D., \& {Jones}, T.~W. 1999, \apjl, 527,
  L113

\bibitem[{{Grenier}(2004)}]{grenier04}
{Grenier}, I.~A. 2004, ArXiv Astrophysics e-prints

\bibitem[{{Grenier} {et~al.}(2005){Grenier}, {Casandjian}, \&
  {Terrier}}]{grenier05}
{Grenier}, I.~A., {Casandjian}, J., \& {Terrier}, R. 2005, Science, 307, 1292

\bibitem[{{Grenier} {et~al.}(1989){Grenier}, {Lebrun}, {Arnaud}, {Dame}, \&
  {Thaddeus}}]{grenier89}
{Grenier}, I.~A., {Lebrun}, F., {Arnaud}, M., {Dame}, T.~M., \& {Thaddeus}, P.
  1989, \apj, 347, 231

\bibitem[{{Hartmann}(2003)}]{hartmann03}
{Hartmann}, L. 2003, \apj, 585, 398

\bibitem[{{Hartmann} {et~al.}(2001){Hartmann}, {Ballesteros-Paredes}, \&
  {Bergin}}]{hartmann01}
{Hartmann}, L., {Ballesteros-Paredes}, J., \& {Bergin}, E.~A. 2001, \apj, 562,
  852

\bibitem[{{Heiles}(1979)}]{heiles79}
{Heiles}, C. 1979, \apj, 229, 533

\bibitem[{{Heiles}(1982)}]{heiles82}
---. 1982, \apj, 262, 135

\bibitem[{{Heiles}(1998)}]{heiles98}
---. 1998, \apj, 498, 689

\bibitem[{{Heiles} {et~al.}(1999){Heiles}, {Haffner}, \& {Reynolds}}]{heiles99}
{Heiles}, C., {Haffner}, L.~M., \& {Reynolds}, R.~J. 1999, in Astronomical
  Society of the Pacific Conference Series, Vol. 168, New Perspectives on the
  Interstellar Medium, ed. A.~R. {Taylor}, T.~L. {Landecker}, \& G.~{Joncas},
  211

\bibitem[{{Heiles} {et~al.}(1996){Heiles}, {Reach}, \& {Koo}}]{heiles96}
{Heiles}, C., {Reach}, W.~T., \& {Koo}, B.-C. 1996, \apj, 466, 191

\bibitem[{{Heiles} \& {Troland}(2003)}]{heiles03}
{Heiles}, C., \& {Troland}, T.~H. 2003, \apj, 586, 1067

\bibitem[{{Heithausen} \& {Thaddeus}(1990)}]{heithausen90}
{Heithausen}, A., \& {Thaddeus}, P. 1990, \apjl, 353, L49

\bibitem[{{Heitsch} \& {Hartmann}(2008)}]{heitsch08c}
{Heitsch}, F., \& {Hartmann}, L. 2008, \apj, 689, 290

\bibitem[{{Heitsch} {et~al.}(2008{\natexlab{a}}){Heitsch}, {Hartmann}, \&
  {Burkert}}]{heitsch08b}
{Heitsch}, F., {Hartmann}, L.~W., \& {Burkert}, A. 2008{\natexlab{a}}, \apj,
  683, 786

\bibitem[{{Heitsch} {et~al.}(2008{\natexlab{b}}){Heitsch}, {Hartmann}, {Slyz},
  {Devriendt}, \& {Burkert}}]{heitsch08a}
{Heitsch}, F., {Hartmann}, L.~W., {Slyz}, A.~D., {Devriendt}, J.~E.~G., \&
  {Burkert}, A. 2008{\natexlab{b}}, \apj, 674, 316

\bibitem[{{Heitsch} {et~al.}(2006){Heitsch}, {Slyz}, {Devriendt}, {Hartmann},
  \& {Burkert}}]{heitsch06}
{Heitsch}, F., {Slyz}, A.~D., {Devriendt}, J.~E.~G., {Hartmann}, L.~W., \&
  {Burkert}, A. 2006, \apj, 648, 1052

\bibitem[{{Heitsch} {et~al.}(2009){Heitsch}, {Stone}, \&
  {Hartmann}}]{heitsch09}
{Heitsch}, F., {Stone}, J.~M., \& {Hartmann}, L.~W. 2009, \apj, 695, 248

\bibitem[{{Hennebelle} {et~al.}(2008){Hennebelle}, {Banerjee},
  {V{\'a}zquez-Semadeni}, {Klessen}, \& {Audit}}]{hennebelle08}
{Hennebelle}, P., {Banerjee}, R., {V{\'a}zquez-Semadeni}, E., {Klessen}, R.~S.,
  \& {Audit}, E. 2008, \aap, 486, L43

\bibitem[{{Hennebelle} \& {P{\'e}rault}(1999)}]{hennebelle99}
{Hennebelle}, P., \& {P{\'e}rault}, M. 1999, \aap, 351, 309

\bibitem[{{Heyer} \& {Terebey}(1998)}]{heyer98}
{Heyer}, M.~H., \& {Terebey}, S. 1998, \apj, 502, 265

\bibitem[{{Heyer} {et~al.}(1996){Heyer}, {Brunt}, {Snell}, {Howe}, {Schloerb},
  {Carpenter}, {Normandeau}, {Taylor}, {Dewdney}, {Cao}, {Terebey}, \&
  {Beichman}}]{heyer96}
{Heyer}, M.~H., {et~al.} 1996, \apjl, 464, L175

\bibitem[{{Hill} {et~al.}(2012){Hill}, {Joung}, {Mac Low}, {Benjamin},
  {Haffner}, {Klingenberg}, \& {Waagan}}]{hill12}
{Hill}, A.~S., {Joung}, M.~R., {Mac Low}, M.-M., {Benjamin}, R.~A., {Haffner},
  L.~M., {Klingenberg}, C., \& {Waagan}, K. 2012, \apj, 750, 104

\bibitem[{{Hollenbach} \& {Salpeter}(1971)}]{hollenbach71}
{Hollenbach}, D., \& {Salpeter}, E.~E. 1971, \apj, 163, 155

\bibitem[{{Hopkins} {et~al.}(2012){Hopkins}, {Quataert}, \&
  {Murray}}]{hopkins12}
{Hopkins}, P.~F., {Quataert}, E., \& {Murray}, N. 2012, \mnras, 421, 3488

\bibitem[{{Hosokawa} \& {Inutsuka}(2006)}]{hosokawa06}
{Hosokawa}, T., \& {Inutsuka}, S.-i. 2006, \apj, 646, 240

\bibitem[{{Inoue} \& {Inutsuka}(2008)}]{inoue08}
{Inoue}, T., \& {Inutsuka}, S. 2008, \apj, 687, 303

\bibitem[{{Inoue} \& {Inutsuka}(2009)}]{inoue09}
---. 2009, \apj, 704, 161

\bibitem[{{Inoue} \& {Inutsuka}(2012)}]{inoue12}
{Inoue}, T., \& {Inutsuka}, S.-i. 2012, \apj, 759, 35

\bibitem[{{Joung} \& {Mac Low}(2006)}]{joung06}
{Joung}, M.~K.~R., \& {Mac Low}, M.-M. 2006, \apj, 653, 1266

\bibitem[{{Joung} {et~al.}(2009){Joung}, {Mac Low}, \& {Bryan}}]{joung09}
{Joung}, M.~R., {Mac Low}, M.-M., \& {Bryan}, G.~L. 2009, \apj, 704, 137

\bibitem[{{Jung} {et~al.}(1996){Jung}, {Koo}, \& {Kang}}]{jung96}
{Jung}, J.~H., {Koo}, B.-C., \& {Kang}, Y.-H. 1996, \aj, 112, 1625

\bibitem[{{Kim} \& {Koo}(2000)}]{kim00}
{Kim}, K.-T., \& {Koo}, B.-C. 2000, \apj, 529, 229

\bibitem[{{Kim} {et~al.}(1999){Kim}, {Dopita}, {Staveley-Smith}, \&
  {Bessell}}]{kim99}
{Kim}, S., {Dopita}, M.~A., {Staveley-Smith}, L., \& {Bessell}, M.~S. 1999,
  \aj, 118, 2797

\bibitem[{{Kim} {et~al.}(2003){Kim}, {Staveley-Smith}, {Dopita}, {Sault},
  {Freeman}, {Lee}, \& {Chu}}]{kim03}
{Kim}, S., {Staveley-Smith}, L., {Dopita}, M.~A., {Sault}, R.~J., {Freeman},
  K.~C., {Lee}, Y., \& {Chu}, Y.-H. 2003, \apjs, 148, 473

\bibitem[{{Kim} \& {Ostriker}(2006)}]{kim06}
{Kim}, W.-T., \& {Ostriker}, E.~C. 2006, \apj, 646, 213

\bibitem[{{Kim} {et~al.}(2002){Kim}, {Ostriker}, \& {Stone}}]{kim02}
{Kim}, W.-T., {Ostriker}, E.~C., \& {Stone}, J.~M. 2002, \apj, 581, 1080

\bibitem[{{Klein} {et~al.}(1994){Klein}, {McKee}, \& {Colella}}]{klein94}
{Klein}, R.~I., {McKee}, C.~F., \& {Colella}, P. 1994, \apj, 420, 213

\bibitem[{{Knee} \& {Brunt}(2001)}]{knee01}
{Knee}, L.~B.~G., \& {Brunt}, C.~M. 2001, \nat, 412, 308

\bibitem[{{Koda} {et~al.}(2009){Koda}, {Scoville}, {Sawada}, {La Vigne},
  {Vogel}, {Potts}, {Carpenter}, {Corder}, {Wright}, {White}, {Zauderer},
  {Patience}, {Sargent}, {Bock}, {Hawkins}, {Hodges}, {Kemball}, {Lamb},
  {Plambeck}, {Pound}, {Scott}, {Teuben}, \& {Woody}}]{koda09}
{Koda}, J., {et~al.} 2009, \apjl, 700, L132

\bibitem[{{Kompaneets}(1960)}]{kompaneets60}
{Kompaneets}, A.~S. 1960, Sov. Phys. Dokl., 5, 46

\bibitem[{{Koo} \& {Heiles}(1988)}]{koo88}
{Koo}, B., \& {Heiles}, C. 1988, in IAU Colloq. 101: Supernova Remnants and the
  Interstellar Medium, ed. {R.~S.~Roger \& T.~L.~Landecker}, 473--+

\bibitem[{{Kun}(1998)}]{kun98}
{Kun}, M. 1998, \apjs, 115, 59

\bibitem[{{Kundt} \& {Mueller}(1987)}]{kundt87}
{Kundt}, W., \& {Mueller}, P. 1987, \apss, 136, 281

\bibitem[{{Lallement} {et~al.}(2003){Lallement}, {Welsh}, {Vergely}, {Crifo},
  \& {Sfeir}}]{lallement03}
{Lallement}, R., {Welsh}, B.~Y., {Vergely}, J.~L., {Crifo}, F., \& {Sfeir}, D.
  2003, \aap, 411, 447

\bibitem[{{Le{\~a}o} {et~al.}(2009){Le{\~a}o}, {de Gouveia Dal Pino},
  {Falceta-Gon{\c c}alves}, {Melioli}, \& {Geraissate}}]{leao09}
{Le{\~a}o}, M.~R.~M., {de Gouveia Dal Pino}, E.~M., {Falceta-Gon{\c c}alves},
  D., {Melioli}, C., \& {Geraissate}, F.~G. 2009, \mnras, 394, 157

\bibitem[{{Loeb} \& {Perna}(1998)}]{loeb98}
{Loeb}, A., \& {Perna}, R. 1998, \apjl, 503, L35+

\bibitem[{{Lombardi} {et~al.}(2008){Lombardi}, {Lada}, \& {Alves}}]{lombardi08}
{Lombardi}, M., {Lada}, C.~J., \& {Alves}, J. 2008, \aap, 489, 143

\bibitem[{{Mac Low} \& {McCray}(1988)}]{maclow88}
{Mac Low}, M., \& {McCray}, R. 1988, \apj, 324, 776

\bibitem[{{Mac Low} {et~al.}(1989){Mac Low}, {McCray}, \& {Norman}}]{maclow89}
{Mac Low}, M., {McCray}, R., \& {Norman}, M.~L. 1989, \apj, 337, 141

\bibitem[{{Mac Low} \& {Glover}(2012)}]{maclow12}
{Mac Low}, M.-M., \& {Glover}, S.~C.~O. 2012, \apj, 746, 135

\bibitem[{{Mac Low} \& {Klessen}(2004)}]{maclow04}
{Mac Low}, M.-M., \& {Klessen}, R.~S. 2004, Reviews of Modern Physics, 76, 125

\bibitem[{{Maciejewski} {et~al.}(1996){Maciejewski}, {Murphy}, {Lockman}, \&
  {Savage}}]{maciejewski96}
{Maciejewski}, W., {Murphy}, E.~M., {Lockman}, F.~J., \& {Savage}, B.~D. 1996,
  \apj, 469, 238

\bibitem[{{Malhotra}(1994{\natexlab{a}})}]{malhotra94b}
{Malhotra}, S. 1994{\natexlab{a}}, \apj, 437, 194

\bibitem[{{Malhotra}(1994{\natexlab{b}})}]{malhotra94a}
---. 1994{\natexlab{b}}, \apj, 433, 687

\bibitem[{{Marx-Zimmer} {et~al.}(2000){Marx-Zimmer}, {Herbstmeier}, {Dickey},
  {Zimmer}, {Staveley-Smith}, \& {Mebold}}]{marx00}
{Marx-Zimmer}, M., {Herbstmeier}, U., {Dickey}, J.~M., {Zimmer}, F.,
  {Staveley-Smith}, L., \& {Mebold}, U. 2000, \aap, 354, 787

\bibitem[{{Mashchenko} \& {Silich}(1994)}]{mashchenko94}
{Mashchenko}, S.~Y., \& {Silich}, S.~A. 1994, {Formation of molecular clouds in
  expanding supershells: 3-D models.}, ed. J.~{Franco}, S.~{Lizano},
  L.~{Aguilar}, \& E.~{Daltabuit}, 202

\bibitem[{{Matsunaga} {et~al.}(2001){Matsunaga}, {Mizuno}, {Moriguchi},
  {Onishi}, {Mizuno}, \& {Fukui}}]{matsunaga01}
{Matsunaga}, K., {Mizuno}, N., {Moriguchi}, Y., {Onishi}, T., {Mizuno}, A., \&
  {Fukui}, Y. 2001, \pasj, 53, 1003

\bibitem[{{McClure-Griffiths} {et~al.}(2002){McClure-Griffiths}, {Dickey},
  {Gaensler}, \& {Green}}]{mcclure02}
{McClure-Griffiths}, N.~M., {Dickey}, J.~M., {Gaensler}, B.~M., \& {Green},
  A.~J. 2002, \apj, 578, 176

\bibitem[{{McClure-Griffiths} {et~al.}(2003){McClure-Griffiths}, {Dickey},
  {Gaensler}, \& {Green}}]{mcclure03}
---. 2003, \apj, 594, 833

\bibitem[{{McClure-Griffiths} {et~al.}(2000){McClure-Griffiths}, {Dickey},
  {Gaensler}, {Green}, {Haynes}, \& {Wieringa}}]{mcclure00}
{McClure-Griffiths}, N.~M., {Dickey}, J.~M., {Gaensler}, B.~M., {Green}, A.~J.,
  {Haynes}, R.~F., \& {Wieringa}, M.~H. 2000, \aj, 119, 2828

\bibitem[{{McCray} \& {Kafatos}(1987)}]{mccray87}
{McCray}, R., \& {Kafatos}, M. 1987, \apj, 317, 190

\bibitem[{{McCray} \& {Snow}(1979)}]{mccray79}
{McCray}, R., \& {Snow}, Jr., T.~P. 1979, \araa, 17, 213

\bibitem[{{McKee} \& {Ostriker}(2007)}]{mckee07}
{McKee}, C.~F., \& {Ostriker}, E.~C. 2007, \araa, 45, 565

\bibitem[{{Meaburn}(1980)}]{meaburn80}
{Meaburn}, J. 1980, \mnras, 192, 365

\bibitem[{{Megeath} {et~al.}(2003){Megeath}, {Biller}, {Dame}, {Leass},
  {Whitaker}, \& {Wilson}}]{megeath03}
{Megeath}, S.~T., {Biller}, B., {Dame}, T.~M., {Leass}, E.~L., {Whitaker}, R.,
  \& {Wilson}, T.~L. 2003, in Revista Mexicana de Astronomia y Astrofisica,
  vol. 27, Vol.~15, Revista Mexicana de Astronomia y Astrofisica Conference
  Series, ed. {J.~Arthur \& W.~J.~Henney}, 151--153

\bibitem[{{Melioli} {et~al.}(2006){Melioli}, {de Gouveia Dal Pino}, {de La
  Reza}, \& {Raga}}]{melioli06}
{Melioli}, C., {de Gouveia Dal Pino}, E.~M., {de La Reza}, R., \& {Raga}, A.
  2006, \mnras, 373, 811

\bibitem[{{Mellema} {et~al.}(2002){Mellema}, {Kurk}, \&
  {R{\"o}ttgering}}]{mellema02}
{Mellema}, G., {Kurk}, J.~D., \& {R{\"o}ttgering}, H.~J.~A. 2002, \aap, 395,
  L13

\bibitem[{{Mer{\'{\i}}n} {et~al.}(2008){Mer{\'{\i}}n}, {J{\o}rgensen},
  {Spezzi}, {Alcal{\'a}}, {Evans}, {Harvey}, {Prusti}, {Chapman}, {Huard}, {van
  Dishoeck}, \& {Comer{\'o}n}}]{merin08}
{Mer{\'{\i}}n}, B., {et~al.} 2008, \apjs, 177, 551

\bibitem[{{Meyerdierks} \& {Heithausen}(1996)}]{meyerdierks96}
{Meyerdierks}, H., \& {Heithausen}, A. 1996, \aap, 313, 929

\bibitem[{{Meyerdierks} {et~al.}(1991){Meyerdierks}, {Heithausen}, \&
  {Reif}}]{meyerdierks91}
{Meyerdierks}, H., {Heithausen}, A., \& {Reif}, K. 1991, \aap, 245, 247

\bibitem[{{Miville-Desch{\^e}nes} {et~al.}(2010){Miville-Desch{\^e}nes},
  {Martin}, {Abergel}, {Bernard}, {Boulanger}, {Lagache}, {Anderson},
  {Andr{\'e}}, {Arab}, {Baluteau}, {Blagrave}, {Bontemps}, {Cohen},
  {Compiegne}, {Cox}, {Dartois}, {Davis}, {Emery}, {Fulton}, {Gry}, {Habart},
  {Huang}, {Joblin}, {Jones}, {Kirk}, {Lim}, {Madden}, {Makiwa}, {Menshchikov},
  {Molinari}, {Moseley}, {Motte}, {Naylor}, {Okumura}, {Pinheiro Gon{\c
  c}alves}, {Polehampton}, {Rod{\'o}n}, {Russeil}, {Saraceno}, {Schneider},
  {Sidher}, {Spencer}, {Swinyard}, {Ward-Thompson}, {White}, \&
  {Zavagno}}]{miville10}
{Miville-Desch{\^e}nes}, M.-A., {et~al.} 2010, \aap, 518, L104

\bibitem[{{Mo{\'o}r} \& {Kiss}(2003)}]{moor03}
{Mo{\'o}r}, A., \& {Kiss}, C. 2003, Commmunications of the Konkoly Observatory
  Hungary, 103, 149

\bibitem[{{Moriguchi} {et~al.}(2002){Moriguchi}, {Onishi}, {Mizuno}, \&
  {Fukui}}]{moriguchi02}
{Moriguchi}, Y., {Onishi}, T., {Mizuno}, A., \& {Fukui}, Y. 2002, in 8th
  Asian-Pacific Regional Meeting, Volume II, ed. {S.~Ikeuchi, J.~Hearnshaw, \&
  T.~Hanawa}, 173--174

\bibitem[{{Moss} {et~al.}(2012){Moss}, {McClure-Griffiths}, {Braun}, {Hill}, \&
  {Madsen}}]{moss12}
{Moss}, V.~A., {McClure-Griffiths}, N.~M., {Braun}, R., {Hill}, A.~S., \&
  {Madsen}, G.~J. 2012, \mnras, 421, 3159

\bibitem[{{Nakamura} {et~al.}(2006){Nakamura}, {McKee}, {Klein}, \&
  {Fisher}}]{nakamura06}
{Nakamura}, F., {McKee}, C.~F., {Klein}, R.~I., \& {Fisher}, R.~T. 2006, \apjs,
  164, 477

\bibitem[{{Norman} \& {Ikeuchi}(1989)}]{norman89}
{Norman}, C.~A., \& {Ikeuchi}, S. 1989, \apj, 345, 372

\bibitem[{{Normandeau} {et~al.}(1996){Normandeau}, {Taylor}, \&
  {Dewdney}}]{normandeau96}
{Normandeau}, M., {Taylor}, A.~R., \& {Dewdney}, P.~E. 1996, \nat, 380, 687

\bibitem[{{Ntormousi} {et~al.}(2011){Ntormousi}, {Burkert}, {Fierlinger}, \&
  {Heitsch}}]{ntormousi11}
{Ntormousi}, E., {Burkert}, A., {Fierlinger}, K., \& {Heitsch}, F. 2011, \apj,
  731, 13

\bibitem[{{Oey} \& {Clarke}(1997)}]{oey97}
{Oey}, M.~S., \& {Clarke}, C.~J. 1997, \mnras, 289, 570

\bibitem[{{Oey} {et~al.}(2001){Oey}, {Clarke}, \& {Massey}}]{oey01}
{Oey}, M.~S., {Clarke}, C.~J., \& {Massey}, P. 2001, in Dwarf galaxies and
  their environment, ed. K.~S. {de Boer}, R.-J. {Dettmar}, \& U.~{Klein}, 181

\bibitem[{{Oey} \& {Garc{\'{\i}}a-Segura}(2004)}]{oey04}
{Oey}, M.~S., \& {Garc{\'{\i}}a-Segura}, G. 2004, \apj, 613, 302

\bibitem[{{Oey} {et~al.}(2005){Oey}, {Watson}, {Kern}, \& {Walth}}]{oey05}
{Oey}, M.~S., {Watson}, A.~M., {Kern}, K., \& {Walth}, G.~L. 2005, \aj, 129,
  393

\bibitem[{{Olano}(1982)}]{olano82}
{Olano}, C.~A. 1982, \aap, 112, 195

\bibitem[{{Olano}(2001)}]{olano01}
---. 2001, \aj, 121, 295

\bibitem[{{Olano} {et~al.}(2006){Olano}, {Meschin}, \& {Niemela}}]{olano06}
{Olano}, C.~A., {Meschin}, P.~I., \& {Niemela}, V.~S. 2006, \mnras, 369, 867

\bibitem[{{Orlando} {et~al.}(2008){Orlando}, {Bocchino}, {Reale}, {Peres}, \&
  {Pagano}}]{orlando08}
{Orlando}, S., {Bocchino}, F., {Reale}, F., {Peres}, G., \& {Pagano}, P. 2008,
  \apj, 678, 274

\bibitem[{{Orlando} {et~al.}(2005){Orlando}, {Peres}, {Reale}, {Bocchino},
  {Rosner}, {Plewa}, \& {Siegel}}]{orlando05}
{Orlando}, S., {Peres}, G., {Reale}, F., {Bocchino}, F., {Rosner}, R., {Plewa},
  T., \& {Siegel}, A. 2005, \aap, 444, 505

\bibitem[{{Patel} {et~al.}(1998){Patel}, {Goldsmith}, {Heyer}, {Snell}, \&
  {Pratap}}]{patel98}
{Patel}, N.~A., {Goldsmith}, P.~F., {Heyer}, M.~H., {Snell}, R.~L., \&
  {Pratap}, P. 1998, \apj, 507, 241

\bibitem[{{Pittard}(2011)}]{pittard11}
{Pittard}, J.~M. 2011, \mnras, 411, L41

\bibitem[{{Pittard} {et~al.}(2009){Pittard}, {Falle}, {Hartquist}, \&
  {Dyson}}]{pittard09}
{Pittard}, J.~M., {Falle}, S.~A.~E.~G., {Hartquist}, T.~W., \& {Dyson}, J.~E.
  2009, \mnras, 394, 1351

\bibitem[{{Pittard} {et~al.}(2010){Pittard}, {Hartquist}, \&
  {Falle}}]{pittard10}
{Pittard}, J.~M., {Hartquist}, T.~W., \& {Falle}, S.~A.~E.~G. 2010, \mnras,
  405, 821

\bibitem[{{Poppel}(1997)}]{poppel97}
{Poppel}, W. 1997, \fcp, 18, 1

\bibitem[{{Preibisch} \& {Zinnecker}(2007)}]{preibisch07}
{Preibisch}, T., \& {Zinnecker}, H. 2007, in IAU Symposium, Vol. 237, IAU
  Symposium, ed. B.~G. {Elmegreen} \& J.~{Palous}, 270--277

\bibitem[{{Reach} {et~al.}(1994){Reach}, {Koo}, \& {Heiles}}]{reach94}
{Reach}, W.~T., {Koo}, B.-C., \& {Heiles}, C. 1994, \apj, 429, 672

\bibitem[{{Reynolds} {et~al.}(2001){Reynolds}, {Sterling}, {Haffner}, \&
  {Tufte}}]{reynolds01}
{Reynolds}, R.~J., {Sterling}, N.~C., {Haffner}, L.~M., \& {Tufte}, S.~L. 2001,
  \apjl, 548, L221

\bibitem[{{Rizzo} \& {Arnal}(1998)}]{rizzo98}
{Rizzo}, J.~R., \& {Arnal}, E.~M. 1998, \aap, 332, 1025

\bibitem[{{Sedov}(1959)}]{sedov59}
{Sedov}, L.~I. 1959, {Similarity and Dimensional Methods in Mechanics}

\bibitem[{{Shetty} \& {Ostriker}(2008)}]{shetty08}
{Shetty}, R., \& {Ostriker}, E.~C. 2008, \apj, 684, 978

\bibitem[{{Shin} {et~al.}(2008){Shin}, {Stone}, \& {Snyder}}]{shin08}
{Shin}, M.-S., {Stone}, J.~M., \& {Snyder}, G.~F. 2008, \apj, 680, 336

\bibitem[{{Tachihara} {et~al.}(2001){Tachihara}, {Toyoda}, {Onishi}, {Mizuno},
  {Fukui}, \& {Neuh{\"a}user}}]{tachihara01}
{Tachihara}, K., {Toyoda}, S., {Onishi}, T., {Mizuno}, A., {Fukui}, Y., \&
  {Neuh{\"a}user}, R. 2001, \pasj, 53, 1081

\bibitem[{{Tasker}(2011)}]{tasker11}
{Tasker}, E.~J. 2011, \apj, 730, 11

\bibitem[{{Tasker} \& {Tan}(2009)}]{tasker09}
{Tasker}, E.~J., \& {Tan}, J.~C. 2009, \apj, 700, 358

\bibitem[{{Taylor} {et~al.}(1999){Taylor}, {Irwin}, {Matthews}, \&
  {Heyer}}]{taylor99}
{Taylor}, A.~R., {Irwin}, J.~A., {Matthews}, H.~E., \& {Heyer}, M.~H. 1999,
  \apj, 513, 339

\bibitem[{{Taylor} {et~al.}(1987){Taylor}, {Dickman}, \& {Scoville}}]{taylor87}
{Taylor}, D.~K., {Dickman}, R.~L., \& {Scoville}, N.~Z. 1987, \apj, 315, 104

\bibitem[{{Tenorio-Tagle}(1981)}]{tenorio81}
{Tenorio-Tagle}, G. 1981, \aap, 94, 338

\bibitem[{{Tenorio-Tagle} \& {Bodenheimer}(1988)}]{tenorio88}
{Tenorio-Tagle}, G., \& {Bodenheimer}, P. 1988, \araa, 26, 145

\bibitem[{{Tenorio-Tagle} \& {Palous}(1987)}]{tenorio87}
{Tenorio-Tagle}, G., \& {Palous}, J. 1987, \aap, 186, 287

\bibitem[{{Tenorio-Tagle} {et~al.}(1990){Tenorio-Tagle}, {Rozyczka}, \&
  {Bodenheimer}}]{tenorio90}
{Tenorio-Tagle}, G., {Rozyczka}, M., \& {Bodenheimer}, P. 1990, \aap, 237, 207

\bibitem[{{Tielens} \& {Hollenbach}(1985)}]{tielens85}
{Tielens}, A.~G.~G.~M., \& {Hollenbach}, D. 1985, \apj, 291, 722

\bibitem[{{Tomisaka}(1992)}]{tomisaka92}
{Tomisaka}, K. 1992, \pasj, 44, 177

\bibitem[{{Tomisaka}(1998)}]{tomisaka98}
---. 1998, \mnras, 298, 797

\bibitem[{{Tomisaka} {et~al.}(1981){Tomisaka}, {Habe}, \&
  {Ikeuchi}}]{tomisaka81}
{Tomisaka}, K., {Habe}, A., \& {Ikeuchi}, S. 1981, \apss, 78, 273

\bibitem[{{Tothill} {et~al.}(2009){Tothill}, {L{\"o}hr}, {Parshley}, {Stark},
  {Lane}, {Harnett}, {Wright}, {Walker}, {Bourke}, \& {Myers}}]{tothill09}
{Tothill}, N.~F.~H., {et~al.} 2009, \apjs, 185, 98

\bibitem[{{van Dishoeck} \& {Black}(1988)}]{vandishoeck88}
{van Dishoeck}, E.~F., \& {Black}, J.~H. 1988, \apj, 334, 771

\bibitem[{{Vanhala} \& {Cameron}(1998)}]{vanhala98}
{Vanhala}, H.~A.~T., \& {Cameron}, A.~G.~W. 1998, \apj, 508, 291

\bibitem[{{V{\'a}zquez-Semadeni}(2010)}]{vazquez10}
{V{\'a}zquez-Semadeni}, E. 2010, in Astronomical Society of the Pacific
  Conference Series, Vol. 438, Astronomical Society of the Pacific Conference
  Series, ed. R.~{Kothes}, T.~L. {Landecker}, \& A.~G. {Willis}, 83

\bibitem[{{V{\'a}zquez-Semadeni} {et~al.}(2007){V{\'a}zquez-Semadeni},
  {G{\'o}mez}, {Jappsen}, {Ballesteros-Paredes}, {Gonz{\'a}lez}, \&
  {Klessen}}]{vazquez07}
{V{\'a}zquez-Semadeni}, E., {G{\'o}mez}, G.~C., {Jappsen}, A.~K.,
  {Ballesteros-Paredes}, J., {Gonz{\'a}lez}, R.~F., \& {Klessen}, R.~S. 2007,
  \apj, 657, 870

\bibitem[{{V{\'a}zquez-Semadeni} {et~al.}(2006){V{\'a}zquez-Semadeni}, {Ryu},
  {Passot}, {Gonz{\'a}lez}, \& {Gazol}}]{vazquez06}
{V{\'a}zquez-Semadeni}, E., {Ryu}, D., {Passot}, T., {Gonz{\'a}lez}, R.~F., \&
  {Gazol}, A. 2006, \apj, 643, 245

\bibitem[{{Wada} \& {Norman}(2001)}]{wada01}
{Wada}, K., \& {Norman}, C.~A. 2001, \apj, 547, 172

\bibitem[{{Wada} {et~al.}(2000){Wada}, {Spaans}, \& {Kim}}]{wada00}
{Wada}, K., {Spaans}, M., \& {Kim}, S. 2000, \apj, 540, 797

\bibitem[{{Walch} {et~al.}(2012){Walch}, {Whitworth}, {Bisbas}, {W{\"u}nsch},
  \& {Hubber}}]{walch12}
{Walch}, S.~K., {Whitworth}, A.~P., {Bisbas}, T., {W{\"u}nsch}, R., \&
  {Hubber}, D. 2012, \mnras, 427, 625

\bibitem[{{Walder} \& {Folini}(1996)}]{walder96}
{Walder}, R., \& {Folini}, D. 1996, \aap, 315, 265

\bibitem[{{Warren} {et~al.}(2011){Warren}, {Weisz}, {Skillman}, {Cannon},
  {Dalcanton}, {Dolphin}, {Kennicutt}, {Koribalski}, {Ott}, {Stilp}, {Van Dyk},
  {Walter}, \& {West}}]{warren11}
{Warren}, S.~R., {et~al.} 2011, \apj, 738, 10

\bibitem[{{Weaver} {et~al.}(1977){Weaver}, {McCray}, {Castor}, {Shapiro}, \&
  {Moore}}]{weaver77}
{Weaver}, R., {McCray}, R., {Castor}, J., {Shapiro}, P., \& {Moore}, R. 1977,
  \apj, 218, 377

\bibitem[{{Weidner} {et~al.}(2010){Weidner}, {Bonnell}, \&
  {Zinnecker}}]{weidner10}
{Weidner}, C., {Bonnell}, I.~A., \& {Zinnecker}, H. 2010, \apj, 724, 1503

\bibitem[{{Weisz} {et~al.}(2009){Weisz}, {Skillman}, {Cannon}, {Dolphin},
  {Kennicutt}, {Lee}, \& {Walter}}]{weisz09}
{Weisz}, D.~R., {Skillman}, E.~D., {Cannon}, J.~M., {Dolphin}, A.~E.,
  {Kennicutt}, Jr., R.~C., {Lee}, J., \& {Walter}, F. 2009, \apj, 704, 1538

\bibitem[{{Wolfire} {et~al.}(2010){Wolfire}, {Hollenbach}, \&
  {McKee}}]{wolfire10}
{Wolfire}, M.~G., {Hollenbach}, D., \& {McKee}, C.~F. 2010, \apj, 716, 1191

\bibitem[{{Wolfire} {et~al.}(1995){Wolfire}, {Hollenbach}, {McKee}, {Tielens},
  \& {Bakes}}]{wolfire95}
{Wolfire}, M.~G., {Hollenbach}, D., {McKee}, C.~F., {Tielens}, A.~G.~G.~M., \&
  {Bakes}, E.~L.~O. 1995, \apj, 443, 152

\bibitem[{{W{\"u}nsch} {et~al.}(2010){W{\"u}nsch}, {Dale}, {Palou{\v s}}, \&
  {Whitworth}}]{wunsch10}
{W{\"u}nsch}, R., {Dale}, J.~E., {Palou{\v s}}, J., \& {Whitworth}, A.~P. 2010,
  \mnras, 407, 1963

\bibitem[{{W{\"u}nsch} {et~al.}(2012){W{\"u}nsch}, {J{\'a}chym}, {Sidorin},
  {Ehlerov{\'a}}, {Palou{\v s}}, {Dale}, {Dawson}, \& {Fukui}}]{wunsch12}
{W{\"u}nsch}, R., {J{\'a}chym}, P., {Sidorin}, V., {Ehlerov{\'a}}, S.,
  {Palou{\v s}}, J., {Dale}, J., {Dawson}, J.~R., \& {Fukui}, Y. 2012, \aap,
  539, A116

\bibitem[{{Yamaguchi} {et~al.}(1999{\natexlab{a}}){Yamaguchi}, {Mizuno},
  {Moriguchi}, {Yonekura}, {Mizuno}, \& {Fukui}}]{yamaguchin99}
{Yamaguchi}, N., {Mizuno}, N., {Moriguchi}, Y., {Yonekura}, Y., {Mizuno}, A.,
  \& {Fukui}, Y. 1999{\natexlab{a}}, \pasj, 51, 765

\bibitem[{{Yamaguchi} {et~al.}(2001{\natexlab{a}}){Yamaguchi}, {Mizuno},
  {Onishi}, {Mizuno}, \& {Fukui}}]{yamaguchi01a}
{Yamaguchi}, R., {Mizuno}, N., {Onishi}, T., {Mizuno}, A., \& {Fukui}, Y.
  2001{\natexlab{a}}, \apjl, 553, L185

\bibitem[{{Yamaguchi} {et~al.}(2001{\natexlab{b}}){Yamaguchi}, {Mizuno},
  {Onishi}, {Mizuno}, \& {Fukui}}]{yamaguchi01b}
---. 2001{\natexlab{b}}, \pasj, 53, 959

\bibitem[{{Yamaguchi} {et~al.}(1999{\natexlab{b}}){Yamaguchi}, {Saito},
  {Mizuno}, {Mine}, {Mizuno}, {Ogawa}, \& {Fukui}}]{yamaguchi99}
{Yamaguchi}, R., {Saito}, H., {Mizuno}, N., {Mine}, Y., {Mizuno}, A., {Ogawa},
  H., \& {Fukui}, Y. 1999{\natexlab{b}}, \pasj, 51, 791

\end{thebibliography}

\end{document}